\documentclass[pra,aps,eqsecnum,amsmath,amssymb,showkeys]{revtex4}
\newcommand{\hh}{{\mathcal{H}}}
\newcommand{\lnp}{{\mathcal{L}}}
\newcommand{\lsp}{{\mathcal{L}}_{+}}
\newcommand{\pen}{\openone}
\newcommand{\Tr}{{\mathrm{Tr}}}
\newcommand{\dmn}{{\mathrm{dim}}}
\newcommand{\bro}{{\boldsymbol{\rho}}}
\newcommand{\vbro}{{\boldsymbol{\varrho}}}
\newcommand{\bsg}{{\boldsymbol{\sigma}}}

\newcommand{\cn}{{\mathsf{C}}}
\newcommand{\gm}{{\mathsf{G}}}

\newcommand{\nm}{{\mathsf{N}}}

\newcommand{\tsm}{{\mathsf{T}}}
\newcommand{\mn}{{\mathsf{M}}}
\newcommand{\um}{{\mathsf{U}}}
\newcommand{\vm}{{\mathsf{V}}}
\newcommand{\ax}{{\mathsf{X}}}
\newcommand{\az}{{\mathsf{Z}}}
\newcommand{\lasf}{{\mathsf{\Lambda}}}
\newcommand{\pisf}{{\mathsf{\Pi}}}
\newcommand{\simf}{{\mathsf{\Sigma}}}
\newcommand{\iu}{{\mathtt{i}}}
\newcommand{\cle}{{\mathcal{E}}}
\newcommand{\clf}{{\mathcal{F}}}
\newcommand{\cli}{{\mathcal{I}}}
\newcommand{\clj}{{\mathcal{J}}}
\newcommand{\clm}{{\mathcal{M}}}
\newcommand{\cln}{{\mathcal{N}}}
\newcommand{\zset}{{\mathbb{C}}}
\newcommand{\mset}{{\mathbb{M}}}

\begin{document}
\clearpage
\preprint{}

\title{R\'{e}nyi and Tsallis formulations of separability conditions in finite dimensions}

\author{Alexey E. Rastegin}
\affiliation{Department of Theoretical Physics, Irkutsk State University,
Gagarin Bv. 20, Irkutsk 664003, Russia}

\begin{abstract}
Separability conditions for a bipartite quantum system of
finite-dimensional subsystems are formulated in terms of R\'{e}nyi
and Tsallis entropies. Entropic uncertainty relations often lead
to entanglement criteria. We propose new approach based on the
convolution of discrete probability distributions. Measurements on
a total system are constructed of local ones according to the
convolution scheme. Separability conditions are derived on the
base of uncertainty relations of the Maassen--Uffink type as well
as majorization relations. On each of subsystems, we use a pair of
sets of subnormalized vectors that form rank-one POVMs. We also
obtain entropic separability conditions for local measurements
with a special structure, such as mutually unbiased bases and
symmetric informationally complete measurements. The relevance of
the derived separability conditions is demonstrated with several
examples. 
\end{abstract}


\maketitle

\pagenumbering{arabic}
\setcounter{page}{1}

\section{Introduction}\label{sec1}

Quantum entanglement stands among fundamentals of the quantum
world. This quantum-mechanical feature was concerned by founders
in the Schr\"{o}dinger ``cat paradox'' paper \cite{cat35} and in
the Einstein--Podolsky--Rosen paper \cite{epr35}. Entanglement is
central to all questions of the emerging technologies of quantum
information. Features of quantum entanglement are currently the
subject of active research (see, e.g., the review \cite{hhhh09}
and references therein). Due to progress in quantum information
processing, both the detection and quantification of entanglement
are very important. In the case of discrete variables, the
positive partial transpose (PPT) criterion \cite{peres96} and the
reduction criterion \cite{mhph1999} are very powerful. On the
other hand, no universal criteria are known even for discrete
variables. Say, the PPT criterion is necessary and sufficient for
$2\times2$ and $2\times3$ systems, but ceases to be so in higher
dimensions \cite{horodecki96}. Separability conditions can be
derived from uncertainty relations of various forms
\cite{guhne04,giovan2004,devic05,guhne06,devic07,huang2010,rastsep}.
The author of \cite{huang2013} proposed a unifying formalism that
reveals links between certain classes of criteria.

Since the Heisenberg principle appeared \cite{heisenberg}, many
formulations and scenarios were studied to understand uncertainty
of complementary observables \cite{hall99,lahti}. One of
approaches to quantifying uncertainty in quantum measurements is
based on the use of entropies \cite{deutsch,kraus,maass}. The
first entropic uncertainty relation for position and momentum was
derived in \cite{hirs} and later improved in
\cite{beckner,birula1}. Entropic uncertainty relations are
currently the subject of active research
\cite{ww10,brud11,cbtw15}. Entropic bounds cannot distinguish the
uncertainty inherent in obtaining a particular selection of the
outcomes \cite{oppwn10}. Fine-grained uncertainty relations were
studied for several scenarios \cite{renf13,rastqip15}.
Majorization approach provides an alternative way to express the
uncertainty principle in terms of probabilities {\sl per se}
\cite{prtv11}. Majorization relations in finite dimensions were
formulated in \cite{prz13,fgg13,rpz14}. The author of \cite{rud15}
derived coarse-grained counterparts of discrete uncertainty
relations based on the concept of majorization. Majorization
uncertainty relations for quantum operations were examined in
\cite{povmkz16}. Majorization-based entropic bounds are sometimes
better than bounds of the Maassen--Uffink type \cite{prz13,rpz14}.

The aim of the present work is to derive separability conditions
on the base of local entropic bounds of various types. To build
total measurement operators, we propose a new unifying scheme
based on the convolution operation. In particular, this approach
allows us to compare majorization uncertainty relations with
relations of the Maassen--Uffink type in the context of their
application in entanglement detection. The paper is organized as
follows. In Sect. \ref{sec2}, we review required material of
matrix analysis and several facts concerning the convolution
operation with discrete indices. In Sect. \ref{sec3}, some known
forms of entropic uncertainty relations are recalled. We also give
a reformulation of the majorization approach for two rank-one
POVMs. Separability conditions in terms of R\'{e}nyi and Tsallis
entropies are formulated in Sect. \ref{sec4}. These conditions
correspond to certain classes of uncertainty relations in
combination with the convolution scheme. In Sect. \ref{sec5}, the
relevance of derived criteria is reasoned with several examples.

\section{Preliminaries}\label{sec2}

In this section, the required definitions will be given. We shall
also prove two preliminary results concerning the convolution of
discrete probability distributions. For two integers $m,n\geq1$,
the symbol $\mset_{m\times{n}}(\zset)$ denotes the space of all
$m\times{n}$ complex matrices. In the case of square matrices,
when $m=n$, we write $\mset_{n}(\zset)$. Each matrix
$\mn\in\mset_{m\times{n}}(\zset)$ can be represented in terms of
the singular value decomposition \cite{hj1990},
\begin{equation}
\mn=\um_{m}\simf\,\um_{n}^{\dagger}
\, , \label{svdx}
\end{equation}
where $\um_{m}\in\mset_{m}(\zset)$ and
$\um_{n}\in\mset_{n}(\zset)$ are unitary. The $m\times{n}$ matrix
$\simf=[[\varsigma_{ij}]]$ has real entries with
$\varsigma_{ij}=0$ for all $i\neq{j}$. If the given matrix $\mn$
has rank $r$, then diagonal entries of $\simf$ can be chosen so
that
\begin{equation}
\varsigma_{11}\geq\cdots\geq\varsigma_{rr}>0=\varsigma_{r+1,r+1}=\cdots=\varsigma_{\ell\ell}
\, , \nonumber
\end{equation}
where $\ell=\min\{m,n\}$. Following \cite{watrous1}, we denote the
Schatten $\infty$-norm as
\begin{equation}
\|\mn\|_{\infty}=\max\bigl\{\varsigma_{jj}(\mn):{\>}1\leq{j}\leq\ell\bigr\}
\, . \label{spnm}
\end{equation}
Treated as an operator norm, the norm (\ref{spnm}) is induced by
the Euclidean norm of vectors \cite{hj1990}. In this sense, it is
sometimes designated with subscript $2$ instead of $\infty$.

The space of linear operators on $d$-dimensional Hilbert space
$\hh$ will be denoted as $\lnp(\hh)$. By $\lsp(\hh)$, we mean the
set of positive operators. The state of a $d$-level
quantum system is described by density matrix $\bro\in\lsp(\hh)$
normalized as $\Tr(\bro)=1$. With respect to the prescribed basis,
vectors of $\hh$ are represented by elements of
$\mset_{d\times{1}}(\zset)$, whereas operators of $\lnp(\hh)$ are
represented by elements of $\mset_{d}(\zset)$.

To formulate schemes for detecting entanglement, we have to build
total-system measurements of local ones. Local measurements are
assumed to be formed by sets of subnormalized vectors. In the
simplest case, we take an orthonormal basis
$\cle=\{|e_{i}\rangle\}$ with $i=0,1,\ldots,d-1$. For the
pre-measurement state $\bro$, $i$-th outcome appears with the
probability $p_{i}(\cle|\bro)=\langle{e}_{i}|\bro|e_{i}\rangle$.
Generalized quantum measurement are typically described in terms
of POVMs. Let $\cln=\{\nm_{i}\}$ be a set of elements of
$\lsp(\hh)$, satisfying the completeness relation
\begin{equation}
\sum_{i=0}^{D-1}\nm_{i}=\pen_{d}
\, . \nonumber
\end{equation}
Such operators form a positive operator-valued measure (POVM). The
probability of $i$-th outcome is expressed as
\begin{equation}
p_{i}(\cln|\bro)=\Tr(\nm_{i}\bro)
\, . \label{pidf}
\end{equation}
In opposite to von Neumann measurements, the number $D$ of
different outcomes can exceed the dimensionality of the Hilbert
space.

Let $p=\{p_{i}\}$ be a probability distribution. For
$0<\alpha\neq1$, the R\'{e}nyi $\alpha$-entropy is defined as
\begin{equation}
R_{\alpha}(p):=\frac{1}{1-\alpha}\>\ln\!\left(\sum\nolimits_{i} p_{i}^{\alpha}\right)
 . \label{rpdf}
\end{equation}
This entropy is a non-increasing function of $\alpha$
\cite{renyi61}. In the limit $\alpha\to1$, we obtain the usual
Shannon entropy
\begin{equation}
H_{1}(p)=-\sum\nolimits_{i} p_{i}\,\ln{p}_{i}
\, . \label{spdf}
\end{equation}
For $\alpha\in(0,1)$, the right-hand side of (\ref{rpdf}) is
certainly concave \cite{ja04}. Convexity properties of R\'{e}nyi's
entropies with orders $\alpha>1$ depend on dimensionality of
probabilistic vectors \cite{bengtsson,ben78}. The binary R\'{e}nyi
entropy is concave for $0<\alpha\leq2$ \cite{ben78}. We also
recall that the R\'{e}nyi entropy is Schur-concave.

Tsallis entropies form another important extension of the Shannon
entropy. For $0<\alpha\neq1$, the Tsallis $\alpha$-entropy is
defined as \cite{tsallis}
\begin{equation}
H_{\alpha}(p):=\frac{1}{1-\alpha}\left(\sum\nolimits_{i} p_{i}^{\alpha}-1\right)
=-\sum\nolimits_{i} p_{i}^{\alpha}\,\ln_{\alpha}(p_{i})
 . \label{tpdf}
\end{equation}
The $\alpha$-logarithm of positive $\xi$ is put here as
$\ln_{\alpha}(\xi)=\bigl(\xi^{1-\alpha}-1\bigr)/(1-\alpha)$. For
$\alpha=1$, Tsallis' $\alpha$-entropy also reduces to
(\ref{spdf}). The right-hand side of (\ref{tpdf}) is a concave
function of probabilities for all $0<\alpha\neq1$. It is
Schur-concave as well. Other properties of R\'{e}nyi and Tsallis
entropies with quantum applications are discussed in
\cite{bengtsson}.

It will be convenient to use norm-like
functionals. For arbitrary $\alpha>0$, we define
\begin{equation}
\|p\|_{\alpha}:=\left(\sum\nolimits_{i} p_{i}^{\alpha}\right)^{1/\alpha}
 . \label{naldf}
\end{equation}
It is a legitimate norm only for $\alpha\geq1$. For
$0<\alpha\neq1$, we then have
\begin{equation}
R_{\alpha}(p)=\frac{\alpha}{1-\alpha}\>\ln\|p\|_{\alpha}
\, . \label{rpdf1}
\end{equation}
By $R_{\alpha}(\cln|\bro)$ and $H_{\alpha}(\cln|\bro)$, we will,
respectively, mean the entropies obtained by substituting the
probabilities (\ref{pidf}) into (\ref{rpdf}) and (\ref{tpdf}).

Let us consider two functions $g=\{g_{i}\}$ and $h=\{h_{i}\}$ of
the discrete variable $i$ that runs $D$ points. Then the
convolution of $g$ and $h$ is introduced as
\begin{equation}
(g*h)_{k}:=\sum_{i=0}^{D-1} g_{i}\,h_{k\ominus{i}}
\, , \label{convid}
\end{equation}
where the sign ``$\ominus$'' denotes the subtraction in
$\mathbb{Z}/D$. The convolution scheme to build measurement
operators is based on a simple but important observation.

\newtheorem{fojen}{Proposition}
\begin{fojen}\label{pon21}
Let $g=\{g_{i}\}$ and $h=\{h_{i}\}$ be positive-valued functions
of integer variable $i\in\{0,1,\ldots,D-1\}$, and let
\begin{equation}
\sum\nolimits_{i} h_{i}=1
\, . \label{norg}
\end{equation}
For $\alpha>1>\beta>0$, we then have
\begin{align}
\|g*h\|_{\alpha}&\leq\|g\|_{\alpha}
\, , \label{fgalp}\\
\|g*h\|_{\beta}&\geq\|g\|_{\beta}
\, . \label{fgbet}
\end{align}
\end{fojen}

{\bf Proof.} For $\alpha>1$, the function $\xi\mapsto\xi^{\alpha}$
has positive second derivative. Due to Jensen's inequality, for
each $k$ we obtain
\begin{equation}
[(g*h)_{k}]^{\alpha}\leq\sum\nolimits_{i} h_{k\ominus{i}}\,g_{i}^{\alpha}
\, . \nonumber
\end{equation}
Summing this with respect to $k$ gives
$\|g*h\|_{\alpha}^{\alpha}\leq\|g\|_{\alpha}^{\alpha}$ due to
(\ref{norg}). This completes the proof of (\ref{fgalp}). The
function $\xi\mapsto\xi^{\beta}$ has negative second derivative
for $0<\beta<1$. Rewriting the above inequalities in opposite
direction, we get the claim (\ref{fgbet}). $\blacksquare$

The notion of majorization is posed as follows. Let us treat
real-valued functions $g=\{g_{i}\}$ and $h=\{h_{i}\}$ of the index
$i\in\{0,1,\ldots,D-1\}$ as $D$-dimensional vectors. The formula
${g}\prec{h}$ implies that, for all $0\leq{k}\leq{D}-1$,
\begin{equation}
\sum_{i=0}^{k} g_{i}^{\downarrow}
\leq\sum_{i=0}^{k} h_{i}^{\downarrow}
\, , \qquad
\sum_{i=0}^{D-1} g_{i}
=\sum_{i=0}^{D-1} h_{i}
\, . \nonumber
\end{equation}
Here, the arrows down imply that the values should be put in the
decreasing order. Our approach to deriving separability conditions
will also use the following lemma.

\newtheorem{macon}[fojen]{Proposition}
\begin{macon}\label{pon22}
Let $p$ and $q$ be two probability distributions supported on the
same finite set; then
\begin{equation}
p*q\prec{p}
\, , \qquad
p*q\prec{q}
\, . \label{pqmaq}
\end{equation}
\end{macon}

{\bf Proof.} It is sufficient to prove only one of the two
relations (\ref{pqmaq}). Let us represent each probability
distribution as a column with $D$ entries. The following result is
well known (see, e.g., theorem II.1.10 of \cite{bhatia97}). The
relation $p*q\prec{p}$ holds if and only if
\begin{equation}
p*q=\tsm p
 \label{th219}
\end{equation}
for some doubly stochastic matrix $\tsm$. A square matrix is
called doubly stochastic, when its entries are positive and the
sum of entries is equal to $1$ in each row and in each column. In
the case considered, the formula (\ref{th219}) directly follows
from the definition of convolution: $i$-th row of $\tsm$ reads
$q_{i\ominus{j}}$, where $j$ runs from $0$ to $D-1$. Hence, each
row of $\tsm$ is obtained by the cyclic shift of the above row by
one step to the right. Thus, each probability $q_{j}$ appears
exactly one time in any row and in any column. Actually, the
matrix $\tsm$ is doubly stochastic. $\blacksquare$

The statement of Proposition \ref{pon22} may be compared
with lemma 1 of \cite{guhne04}. Here, we prefer to give another
formulation with emphasizing the role of convolution. In addition,
our scheme is rather formulated in terms of measurement operators
including the case of POVMs. As lemma 1 of \cite{guhne04} deals
with observables, it does not seem to be applicable immediately in
these settings.

Let $\cle=\bigl\{|e_{i}\rangle\bigr\}$ and
$\cle^{\prime}=\bigl\{|e_{j}^{\prime}\rangle\bigr\}$ be two
orthonormal bases in a $d$-dimensional Hilbert space $\hh$. They
are said to be mutually unbiased if and only if for all $i$ and
$j$,
\begin{equation}
\bigl|\langle{e}_{i}|e_{j}^{\prime}\rangle\bigr|=\frac{1}{\sqrt{d}}
\ . \label{twbs}
\end{equation}
Several orthonormal bases form a set of mutually unbiased bases
(MUBs), when them are pairwise mutually unbiased. Such bases have
found use in many questions of quantum information theory (see
\cite{bz10} and references therein). When $d$ is a prime power, we
can certainly construct $d+1$ MUBs \cite{bz10}. It is
based on properties of prime powers and corresponding finite field
\cite{wf89,kr04}.

There exist measurements such that each of them uniquely determine
every possible state by the measurement statistics that it alone
generates. Measurements with this property are said to be
informationally complete \cite{watrous1}. Symmetric
informationally complete measurements have a symmetric structure
in their elements. In $d$-dimensional Hilbert space, we consider a
set of $d^{2}$ rank-one operators of the form
\begin{equation}
|f_{i}\rangle\langle{f}_{i}|=\frac{1}{d}
\,|\phi_{i}\rangle\langle\phi_{i}|
\ . \label{usic}
\end{equation}
If the normalized vectors $|\phi_{j}\rangle$ all satisfy the
condition
\begin{equation}
\bigl|\langle\phi_{i}|\phi_{j}\rangle\bigr|^{2}=\frac{1}{d+1}
\qquad (i\neq{j})
\ , \label{undn1}
\end{equation}
the set of operators (\ref{usic}) is a symmetric informationally
complete POVM (SIC-POVM) \cite{rbsc04}. It was conjectured that
SIC-POVMs exist in all dimensions \cite{appl2005}. The existence
of SIC-POVMs has been shown analytically or numerically for all
dimensions up to 67 \cite{grassl10}. Connections
between MUBs and SIC-POVMs are discussed in \cite{adf07,ruskai09}.

Since basic constructions of MUBs are related to prime power $d$,
one can try to get an appropriate modification. The authors of
\cite{kagour} proposed the concept of mutually unbiased
measurements (MUMs). Using weaker requirements, a complete set of
$d+1$ MUMs exists for all $d$. Let us consider two POVM
measurements $\cln=\{\nm_{i}\}$ and
$\cln^{\prime}=\{\nm_{j}^{\prime}\}$. Each of them contains $d$
elements such that
\begin{align}
& \Tr(\nm_{i})=\Tr(\nm_{j}^{\prime})=1
\ , \label{tmn1}\\
& \Tr(\nm_{i}\nm_{j}^{\prime})=\frac{1}{d}
\ . \label{dmn1}
\end{align}
The POVM elements are all of trace one, but generally not of rank
one. The formula (\ref{dmn1}) is used instead of the squared
formula (\ref{twbs}). Two different elements of the same POVM
$\cln$ satisfy
\begin{equation}
\Tr(\nm_{i}\nm_{j})=\delta_{ij}{\,}\varkappa
+(1-\delta_{ij}){\>}\frac{1-\varkappa}{d-1}
\ , \label{mjmk}
\end{equation}
where $\varkappa$ is the efficiency parameter \cite{kagour}. The
same condition is imposed on the elements of $\cln^{\prime}$. By
$\varkappa$, we characterize how close the POVM elements are to
rank-one projectors \cite{kagour}. In general, one satisfies
\cite{kagour}
\begin{equation}
\frac{1}{d}<\varkappa\leq1
\ . \nonumber
\end{equation}
For $\varkappa=1/d$ we have the trivial case, in which
$\nm_{i}=\pen_{d}/d$ for all $i$. The value $\varkappa=1$, if
possible, gives the standard case of mutually unbiased bases. In
principle, we can only say that the maximal efficiency can be
reached for prime power $d$. More precise bounds on $\varkappa$
depend on an explicit construction of POVM elements \cite{kagour}.

Similar ideas can be used to generalize SIC-POVMs. For all finite
$d$, a common construction has been given \cite{kgour13}. Consider
a POVM with $d^{2}$ elements $\nm_{i}$, which satisfy the
following two conditions. First, for all $i=0,\ldots,d^{2}-1$ we
have
\begin{equation}
\Tr(\nm_{i}\nm_{i})=a
\, . \label{ficnd}
\end{equation}
Second, the pairwise inner products are all symmetric, namely
\begin{equation}
\Tr(\nm_{i}\nm_{j})=b
\qquad (i\neq{j})
\, . \label{secnd}
\end{equation}
Then, the operators $\nm_{i}$ form a general SIC-POVM. Combining
the conditions (\ref{ficnd}) and (\ref{secnd}) with the
completeness relation finally gives \cite{kgour13}
\begin{equation}
b=\frac{1-ad}{d(d^{2}-1)}
\ . \label{bvia}
\end{equation}
We also get $\Tr(\nm_{i})=1/d$ for all $i$. A deviation of general
SIC-POVM from the usual one is completely characterized by $a$. In
general, this parameter is restricted as \cite{kgour13}
\begin{equation}
\frac{1}{d^{3}}<{a}\leq\frac{1}{d^{2}}
\ . \nonumber
\end{equation}
The value $a=1/d^{3}$ gives $\nm_{i}=\pen_{d}/d^{2}$, so that the
measurement is not informationally complete. The value $a=1/d^{2}$
is achieved, when the POVM elements are all rank-one
\cite{kgour13}. The latter is actually the case of usual
SIC-POVMs, when POVM elements appear due to (\ref{usic}). Even if
usual SIC-POVMs exist in all dimensions, they are rather hard to
construct. General SIC-POVMs have a similar structure that makes
them appropriate in determining an informational content of a
quantum state.

\section{Some forms of uncertainty relations}\label{sec3}

In this section, we recall some of existing formulations
of the uncertainty principle. We begin with uncertainty relations of the
Maassen--Uffink type \cite{maass}. Then, majorization
uncertainty relations of the papers \cite{prz13,rpz14} will be
applied. Uncertainty relations for measurements with a special
structure should also be considered.

The used orthonormal bases are denoted by $\cle=\bigl\{|e_{i}\rangle\bigr\}$
and $\cle^{\prime}=\bigl\{|e_{j}^{\prime}\rangle\bigr\}$ with
$i,j=0,\ldots,d-1$. If the pre-measurement state is described by
normalized density matrix $\bro\in\lsp(\hh)$, then the
corresponding probabilities appear as $p_{i}=p_{i}(\cle|\bro)$ and
$q_{j}=p_{j}(\cle^{\prime}|\bro)$. Entropic uncertainty relations
of the Maassen--Uffink type were inspired by Kraus \cite{kraus}
and later proved in \cite{maass}. To the orthonormal bases
$\cle=\bigl\{|e_{i}\rangle\bigr\}$ and
$\cle^{\prime}=\bigl\{|e_{j}^{\prime}\rangle\bigr\}$, we assign
\begin{equation}
\eta(\cle,\cle^{\prime}):=
\max\Bigl\{\bigl|\langle{e}_{i}|e_{j}^{\prime}\rangle\bigr|:{\>}0\leq{i},j\leq{d}-1\Bigr\}
\, . \label{mupr}
\end{equation}
Due to Riesz's theorem \cite{riesz27}, we have
\begin{align}
\|p\|_{\alpha}&\leq\eta^{2(1-\beta)/\beta}\,\|q\|_{\beta}
\, , \label{cmu1}\\
\|q\|_{\alpha}&\leq\eta^{2(1-\beta)/\beta}\,\|p\|_{\beta}
\, , \label{cmu2}
\end{align}
where $1/\alpha+1/\beta=2$ and $\alpha>1>\beta$. Hence, various
uncertainty relations in terms of generalized entropies can be
derived. For some reasons, we will begin a derivation of
separability conditions just with (\ref{cmu1}) and (\ref{cmu2}).

The formulas (\ref{cmu1}) and (\ref{cmu2}) are immediately
generalized to POVM measurements \cite{rast104}. Here, we restrict
a consideration to especially important case of rank-one POVMs.
Let $\clf=\bigl\{|f_{i}\rangle\bigr\}$ and
$\clf^{\prime}=\bigl\{|f_{j}^{\prime}\rangle\bigr\}$ be two sets
of $D$ subnormalized vectors such that
\begin{equation}
\sum_{i=0}^{D-1} |f_{i}\rangle\langle{f}_{i}|=\pen_{d}
\, , \qquad
\sum_{j=0}^{D-1} |f_{j}^{\prime}\rangle\langle{f}^{\prime}_{j}|=\pen_{d}
\, . \label{abpen}
\end{equation}
Here, we typically deal with $D>d$. The author of \cite{hall97}
pointed out haw the Maassen--Uffink relation is generalized to
such measurements. Following \cite{rastmub}, we are rather
interested in extending just (\ref{cmu1}) and (\ref{cmu2}).
Replacing (\ref{mupr}) with
\begin{equation}
\eta(\clf,\clf^{\prime}):=
\max\Bigl\{\bigl|\langle{f}_{i}|f_{j}^{\prime}\rangle\bigr|:{\>}0\leq{i},j\leq{D}-1\Bigr\}
\, , \label{mupr1}
\end{equation}
the relations (\ref{cmu1}) and (\ref{cmu2}) hold under the
conditions $1/\alpha+1/\beta=2$ and $\alpha>1>\beta$.

We now recall the majorization approach to uncertainty relations
in finite dimensions. Applications of this approach beyond this
case are discussed in \cite{prtv11,rud15}. Let $p$ and $q$
denote the probabilistic vectors generated by two quantum
measurements in the same prepared state. The basic idea is to
majorize some binary combination of $p$ and $q$ by a third vector
with bounding elements. To do so, the authors of
\cite{prz13,rpz14} inspected norms of submatrices of a certain
unitary matrix.

To the orthonormal bases $\cle=\bigl\{|e_{i}\rangle\bigr\}$ and
$\cle^{\prime}=\bigl\{|e_{j}^{\prime}\rangle\bigr\}$, one assigns the
unitary $d\times{d}$ matrix $\vm(\cle,\cle^{\prime})$ with entries
$v_{ij}=\langle{e}_{i}|e_{j}^{\prime}\rangle$. By
$\mathcal{SUB}(\vm,k)$, we mean the set of all its submatrices of
class $k$ defined by
\begin{equation}
{\mathcal{SUB}}(\vm,k):=
\bigl\{
\mn\in\mset_{r\times{r}^{\prime}}(\zset):{\>}r+r^{\prime}=k+1,{\>}
\mn {\text{ is a submatrix of }} \vm
\bigr\}
\, . \label{subvk}
\end{equation}
The positive integer $k$ runs all the values allowed by the
condition $r+r^{\prime}=k+1$. The majorization relations of
\cite{prz13,rpz14} are expressed in terms of quantities
\begin{equation}
s_{k}:=\max\bigl\{
\|\mn\|_{\infty}:{\>}\mn\in\mathcal{SUB}(\vm,k)
\bigr\}
\, . \label{skdf}
\end{equation}
It will be convenient to label these quantities by integers
starting with $1$. Due to completeness and orthonormality of each
bases, one has $s_{d}=1$ and, therefore, $s_{k}=1$ for all
$d\leq{k}\leq2d-1$.

The authors of \cite{rpz14} proved the majorization relation
\begin{align}
&p\oplus{q}\prec\{1\}\oplus{w}
\, , \label{wdd}\\
&w=(s_{1},s_{2}-s_{1},\ldots,s_{d}-s_{d-1})
\, . \label{wdd1}
\end{align}
This majorizing vector is completed by $s_{d}-s_{d-1}$, since
$s_{k}=1$ for $d\leq{k}\leq2d-1$ and further differences are all
zero. The following entropic bounds follow from (\ref{wdd}). For
$0<\alpha\leq1$, it holds that
\begin{equation}
R_{\alpha}(p)+R_{\alpha}(q)\geq{R}_{\alpha}(w)
\, . \label{nwmr0}
\end{equation}
For $\alpha>1$, the sum of two R\'{e}nyi entropies obeys another
inequality \cite{rpz14}
\begin{equation}
R_{\alpha}(p)+R_{\alpha}(q)\geq
\frac{2}{1-\alpha}{\>}
\ln\!\left(
\frac{1+\|w\|_{\alpha}^{\alpha}}{2}
\right)
 . \label{nwmr1}
\end{equation}
Majorization relations of the tensor-product type were first
considered in \cite{prz13,fgg13}. The authors of \cite{prz13}
showed that
\begin{equation}
p\otimes{q}\prec{w}^{\prime}
\, , \label{pq0w}
\end{equation}
where the majorizing vector is put as
\begin{equation}
w^{\prime}=(t_{1},t_{2}-t_{1},\ldots,t_{d}-t_{d-1})
\, , \qquad
t_{k}=\frac{(1+s_{k})^{2}}{4}
\, . \label{wpdd}
\end{equation}
Combining (\ref{pq0w}) with Schur-concavity of the R\'{e}nyi
entropy, for $\alpha>0$ we have \cite{prz13,fgg13}
\begin{equation}
R_{\alpha}(p)+R_{\alpha}(q)\geq{R}_{\alpha}(w^{\prime})
\, . \label{oldmr}
\end{equation}
For $0<\alpha\leq1$, we will choose (\ref{nwmr0}), since
$\omega\prec\omega^{\prime}$ and
$R_{\alpha}(\omega)\geq{R}_{\alpha}(\omega^{\prime})$
\cite{rpz14}. Nevertheless, the relation (\ref{oldmr}) is
useful for $\alpha>1$. The sum of two Tsallis $\alpha$-entropies
is bounded from below similarly to (\ref{nwmr0}). For any
$\alpha>0$ we have \cite{rpz14}
\begin{equation}
H_{\alpha}(p)+H_{\alpha}(q)\geq{H}_{\alpha}(w)
\, . \label{nwmr01}
\end{equation}

As we plan to deal with rank-one POVMs, the majorization approach
should be reformulated appropriately. In principle, a general way
of extension was considered in \cite{povmkz16}. However, that
paper focus on quantum operations described in terms of Kraus
operators. Rank-one POVMs are so important that we prefer to give
an explicit derivation. In this case, the majorization approach is
based on the following statement.

\newtheorem{abmaj}[fojen]{Proposition}
\begin{abmaj}\label{pon31}
Let each of sets $\clf=\bigl\{|f_{i}\rangle\bigr\}$ and
$\clf^{\prime}=\bigl\{|f_{j}^{\prime}\rangle\bigr\}$ contain $D$
subnormalized vectors that form rank-one POVM in $d$-dimensional
space $\hh$. Let $\cli$ and $\clj$ be two subsets of the set
$\{0,\ldots,D-1\}$. For arbitrary density matrix $\bro$, we have
\begin{equation}
\sum\nolimits_{i\in\cli}p_{i}(\clf|\bro)+
\sum\nolimits_{j\in\clj}p_{j}(\clf^{\prime}|\bro)\leq
1+\|\cn_{\cli}\cn_{\clj}^{\dagger}\|_{\infty}
\ . \label{rlmij}
\end{equation}
Here, the $|\cli|\times{d}$ matrix $\cn_{\cli}$ is formed by rows
$\langle{f}_{i}|$ with $i\in\cli$, and the $|\clj|\times{d}$
matrix $\cn_{\clj}$ is formed by rows $\langle{f}_{j}^{\prime}|$
with $j\in\clj$.
\end{abmaj}

{\bf Proof.} For definiteness, we write
$\cli=\{i_{1},\ldots,i_{m}\}$ and $\clj=\{j_{1},\ldots,j_{n}\}$,
whence
\begin{equation}
\cn_{\cli}=
\begin{pmatrix}
\langle{f}_{i_{1}}| \\
\cdots \\
\langle{f}_{i_{m}}|
\end{pmatrix}
 , \qquad
\cn_{\clj}=
\begin{pmatrix}
\langle{f}^{\prime}_{j_{1}}| \\
\cdots \\
\langle{f}^{\prime}_{j_{n}}|
\end{pmatrix}
 . \nonumber
\end{equation}
It will be sufficient to prove the claim (\ref{rlmij}) for pure
states. Its validity for mixed states follows by the spectral
decomposition. Keeping in mind matrix relations of the form
\begin{equation}
\cn_{\cli}^{\dagger}\cn_{\cli}=
\sum\nolimits_{i\in\cli} |f_{i}\rangle\langle{f}_{i}|
\, , \label{claar}
\end{equation}
we have
\begin{equation}
\sum\nolimits_{i\in\cli} p_{i}(\clf|\psi)+
\sum\nolimits_{j\in\clj} p_{j}(\clf^{\prime}|\psi)=
\langle\psi|\gm^{\dagger}\gm|\psi\rangle
\, , \qquad
\gm=
\begin{pmatrix}
\cn_{\cli} \\
\cn_{\clj}
\end{pmatrix}
 . \nonumber
\end{equation}
Due to properties of the spectral norm, we obtain
\begin{equation}
\langle\psi|\gm^{\dagger}\gm|\psi\rangle\leq\|\gm^{\dagger}\gm\|_{\infty}
=\|\gm\gm^{\dagger}\|_{\infty}
\leq\max\bigl\{
\|\cn_{\cli}\|_{\infty}^{2},\|\cn_{\clj}\|_{\infty}^{2}
\bigr\}+\|\cn_{\cli}\cn_{\clj}^{\dagger}\|_{\infty}
\, . \label{cacb0}
\end{equation}
The justification of (\ref{cacb0}) is very similar to the proof of
proposition 2 of \cite{povmkz16}. The definition of $\cn_{\cli}$
and $\cn_{\clj}$ is the only distinction. Let us put the complete
$D\times{d}$ matrices such that $\cn_{\clf}$ is formed by all the
rows $\langle{f}_{i}|$, and $\cn_{\clf^{\prime}}$ is formed by all
the rows $\langle{f}_{j}^{\prime}|$. By submultiplicativity of the
spectral norm, one gets
\begin{equation}
\|\cn_{\cli}\|_{\infty}\leq\|\cn_{\clf}\|_{\infty}
\, , \qquad
\|\cn_{\clj}\|_{\infty}\leq\|\cn_{\clf^{\prime}}\|_{\infty}
\, . \label{subab}
\end{equation}
It follows from (\ref{abpen}) that
$\cn_{\clf}^{\dagger}\cn_{\clf}=\cn_{\clf^{\prime}}^{\dagger}\cn_{\clf^{\prime}}=\pen_{d}$,
whence $\|\cn_{\clf}\|_{\infty}=\|\cn_{\clf^{\prime}}\|_{\infty}=1$.
Combining the latter with (\ref{cacb0}) and (\ref{subab})
completes the proof. $\blacksquare$

Using (\ref{rlmij}), we can now extend (\ref{nwmr0}),
(\ref{nwmr1}), (\ref{oldmr}), and (\ref{nwmr01}). Denoting
$p=p(\clf|\bro)$ and $q=p(\clf^{\prime}|\bro)$, these uncertainty
relations are all valid with the following changes. The quantities
(\ref{skdf}) are now calculated with $1\leq{k}\leq2D-1$ for the
$D\times{D}$ matrix
\begin{equation}
\vm(\clf,\clf^{\prime})=\bigl[\bigl[\langle{f}_{i}|f_{j}^{\prime}\rangle\bigr]\bigr]
\, . \label{neww}
\end{equation}
Combining
$\cn_{\clf^{\prime}}^{\dagger}\cn_{\clf^{\prime}}=\pen_{d}$ with
$\vm=\cn_{\clf}\cn_{\clf^{\prime}}^{\dagger}$ leads to
$\vm\vm^{\dagger}=\cn_{\clf}\cn_{\clf}^{\dagger}$ and
$\|\vm\|_{\infty}=1$. Thus, we certainly have $s_{k}=1$ for
$k=2D-1$. Let $D_{\star}$ denote the first index with the property
$s_{D_{\star}}=1$. Since vectors of the sets $\clf$ and
$\clf^{\prime}$ are all subnormalized, we will have
$D\leq{D}_{\star}\leq2D-1$.

The vectors (\ref{wdd1}) and (\ref{wpdd}) should then include
differences up to $s_{D_{\star}}-s_{D_{\star}-1}$ and
$t_{D_{\star}}-t_{D_{\star}-1}$, respectively.  In the following,
the uncertainty relations of this section will be applied in
deriving entanglement criteria. Among majorization-based
relations, we will mainly use (\ref{nwmr0}) and (\ref{nwmr01})
with respect to the values of $\alpha$, for which the
corresponding entropy is certainly concave.

It is often expedient to check entanglement with specially
designed measurements. For example, mutually unbiased measurements
are treated to be capable for such purposes. Another interesting
way is connected with symmetric informationally complete
measurements. Here, we will use entropic uncertainty relations
derived in \cite{rastmub,remum15}.

Let $\bigl\{\cle^{(1)},\ldots,\cle^{(K)}\bigr\}$ be a set of $K$
MUBs in $d$-dimensional space $\hh$. For $\alpha\in(0;2]$, the sum
of R\'{e}nyi's entropies obeys the state-independent bound \cite{rastmub}
\begin{equation}
\frac{1}{K}\,\sum_{t=1}^{K} R_{\alpha}(\cle^{(t)}|\bro)\geq
\ln\!\left(\frac{Kd}{d+K-1}\right)
 . \label{rmub02}
\end{equation}
Note that the right-hand side of (\ref{rmub02}) is independent of
$\alpha$, whereas R\'{e}nyi's entropy does not increase with
growth of $\alpha$. To obtain more sensitive criteria, we should
take largest orders providing concavity of entropies. Therefore,
we will use (\ref{rmub02}) with $\alpha=1$ for arbitrary $d$ and
with $\alpha=2$ for $d=2$. For $\alpha\in(0;2]$ and arbitrary
state $\bro$ on $\hh$, the sum of Tsallis' entropies satisfies the
state-independent bound
\begin{equation}
\frac{1}{K}\,\sum_{t=1}^{K} H_{\alpha}(\cle^{(t)}|\bro)
\geq\ln_{\alpha}\!\left(\frac{Kd}{d+K-1}\right)
 . \label{tmub02}
\end{equation}
The results (\ref{rmub02}) and (\ref{tmub02}) are based on the
inequality
\begin{equation}
\sum_{t=1}^{K} \sum_{i=0}^{d-1} p_{i}(\cle^{(t)}|\bro)^{2}
\leq\Tr(\bro^{2})+\frac{K-1}{d}
\leq1+\frac{K-1}{d}
\ , \label{indk}
\end{equation}
derived in \cite{molm09}. Of course, the existence of $K$ MUBs
should be proved independently. We will study entropic formulation
of entanglement criteria based on (\ref{indk}). It differs from
the previous approach considered in \cite{shbah12}. Using
(\ref{indk}), the authors of \cite{shbah12} put a specific
correlation measure that is bounded from above for separable
states.

For mutually unbiased measurements, the following extension of
(\ref{indk}) takes place \cite{remum15}. Let
$\bigl\{\cln^{(1)},\ldots,\cln^{(K)}\bigr\}$ be a set of $K$ MUMs
of the efficiency $\varkappa$. For arbitrary $\bro$, we then have
\cite{remum15}
\begin{equation}
\sum_{t=1}^{K} \sum_{i=0}^{d-1} p_{i}(\cln^{(t)}|\bro)^{2}
\leq
\frac{1-\varkappa+(\varkappa{d}-1){\,}\Tr(\bro^{2})}{d-1}
+\frac{K-1}{d}
\leq
\varkappa+\frac{K-1}{d}
\ . \label{indkm2}
\end{equation}
For $d+1$ MUMs, the inequality (\ref{indkm2}) is actually
saturated \cite{remum15}. For pure states, this result was shown
in \cite{kagour} and then applied for entanglement detection in
\cite{fei14}. For $\alpha\in(0;2]$, the inequality (\ref{indkm2}) gives
\begin{align}
\frac{1}{K}\,\sum_{t=1}^{K} R_{\alpha}(\cln^{(t)}|\bro)&\geq
\ln\!\left(\frac{Kd}{\varkappa{d}+K-1}\right)
 , \label{rmum02}\\
\frac{1}{K}\,\sum_{t=1}^{K} H_{\alpha}(\cln^{(t)}|\bro)&\geq
\ln_{\alpha}\!\left(\frac{Kd}{\varkappa{d}+K-1}\right)
 . \label{tmum02}
\end{align}
The existence of $K$ MUMs of some efficiency was proved up to
$K=d+1$ \cite{kagour}.

For a SIC-POVM $\clf=\{|f_{i}\rangle\}$, the index of coincidence
can also be calculated exactly. For the pre-measurement state
$\bro$, it holds that \cite{rastmub}
\begin{equation}
\sum_{i=0}^{d^{2}-1} p_{i}(\clf|\bro)^{2}=\frac{\Tr(\bro^{2})+1}{d(d+1)}
\leq\frac{2}{d(d+1)}
\ . \label{indc1}
\end{equation}
As was briefly noticed in \cite{rastmub}, this result allows to
build a SIC-POVM scheme for entanglement detection. The following
uncertainty relations were derived due to (\ref{indc1}). For
$\alpha\in(0;2]$ and any density matrix $\bro$ on $\hh$, we have
\cite{rastmub}
\begin{align}
R_{\alpha}(\clf|\bro)&\geq
\ln\!\left(\frac{d(d+1)}{2}\right)
 , \label{rsic02}\\
H_{\alpha}(\clf|\bro)&\geq
\ln_{\alpha}\!\left(\frac{d(d+1)}{2}\right)
 . \label{tsic02}
\end{align}

Let general SIC-POVM $\cln=\{\nm_{i}\}$ be characterized by the
parameter $a$ in the sense of (\ref{ficnd}). For the given
pre-measurement state $\bro$, we have \cite{rastsic}
\begin{equation}
\sum_{i=0}^{d^{2}-1} p_{i}(\cln|\bro)^{2}
=\frac{(ad^{3}-1){\,}\Tr(\bro^{2})+d(1-ad)}{d(d^{2}-1)}
\leq\frac{ad^{2}+1}{d(d+1)}
\ . \label{indc2}
\end{equation}
For a usual SIC-POVM, when $a=d^{-2}$, the result (\ref{indc2}) is
reduced to (\ref{indc1}). For $\alpha\in(0;2]$ and arbitrary
density matrix $\bro$ on $\hh$, one gets \cite{rastsic}
\begin{align}
R_{\alpha}(\cln|\bro)&\geq
\ln\!\left(\frac{d(d+1)}{ad^{2}+1}\right)
 , \label{rgsic02}\\
H_{\alpha}(\cln|\bro)&\geq
\ln_{\alpha}\!\left(\frac{d(d+1)}{ad^{2}+1}\right)
 . \label{tgsic02}
\end{align}
Due to (\ref{indc2}), general SIC-POVMs can be used
for entanglement detection. The authors of \cite{fei15} gave an
appropriate form of correlation measures proposed in
\cite{shbah12} and reformulated for usual SIC-POVMs in
\cite{rastmub}.

\section{Formulation of separability conditions}\label{sec4}

This section is devoted to separability conditions that follow
from uncertainty relations listed in the previous section. In
particular, we will present separability conditions on the base of
majorization uncertainty relations. Used uncertainty relations are
local in the sense that they are posed for one of subsystems. A
utility of entanglement criteria based on local uncertainty
relations was justified in \cite{guhne06}. To formulate
separability conditions, some definitions should be recalled.

We consider a bipartite system of $d$-level subsystems $A$ and
$B$. The tensor product $\hh_{AB}=\hh_{A}\otimes\hh_{B}$ is the
total Hilbert space. Any state of the total system is described by
density matrix $\bro_{AB}\in\lsp(\hh_{AB})$. Product states
written as $\bro_{A}\otimes\bro_{B}$ reveal no correlations
between subsystems. A bipartite mixed state is called separable,
when its density matrix can be represented as a convex combination
of product states \cite{werner89,zhsl98,dcls2017}. For more formal
results about separable operators and states, see chapter 6 of
\cite{watrous1}.

Let us proceed to building measurements on a bipartite system
according to the convolution scheme. Its advance is that POVM
measurements are naturally treated in the context of entanglement
detection. In a certain sense, this scheme is a genuine
development of the approach studied in \cite{rastsep}. Total
measurement operators are constructed as follows.

\newtheorem{defsc}{Definition}
\begin{defsc}\label{def1}
Let $\cln_{A}=\{\nm_{Ai}\}$ and $\cln_{B}=\{\nm_{Bj}\}$ be
$D$-outcome POVMs in $\hh_{A}$ and $\hh_{B}$, respectively. We
call a POVM $\clm(\cln_{A},\cln_{B})=\{\pisf_{k}\}$ to be
constructed according to the convolution scheme, when
\begin{equation}
\pisf_{k}:=
\sum_{i=0}^{D-1}{\nm_{Ai}\otimes\nm_{Bk\ominus{i}}}
\, . \label{pisfdc}
\end{equation}
Here, the sign ``$\ominus$'' denotes the subtraction in
$\mathbb{Z}/D$ and $k\in\{0,1,\ldots,D-1\}$.
\end{defsc}

For each of two subsystems, we will use several measurements
marked by the label $t$. For each $\clm^{(t)}$ built according to
Definition \ref{def1}, one has
\begin{equation}
p(\clm^{(t)}|\bro_{A}\otimes\bro_{B})=
p(\cln_{A}^{(t)}|\bro_{A})*p(\cln_{B}^{(t)}|\bro_{B})
\, . \label{pppcn1}
\end{equation}
For product states, each resolution $\clm^{(t)}$ generates the
convolution of two distributions assigned to local measurements.
Together with entropic bounds, this fact allows us to get
inequalities that are satisfied by any convex combination of
product states. The convolution operation is also important in
deriving entropic entanglement criteria for a bipartite system
with continuous-variables \cite{wtstd09,stw11}. They have recently
been extended to multipartite systems \cite{rastpra17}. Our first
result is posed as follows.

\newtheorem{musep}[fojen]{Proposition}
\begin{musep}\label{pon41}
Let each of sets $\clf_{A}^{(1)}$ and $\clf_{A}^{(2)}$ of
subnormalized vectors form rank-one POVM in $\hh_{A}$, and let
each of sets $\clf_{B}^{(1)}$ and $\clf_{B}^{(2)}$ of
subnormalized vectors form rank-one POVM in $\hh_{B}$. Let two
POVMs $\clm^{(t)}(\clf_{A}^{(t)},\clf_{B}^{(t)})$ in
$\hh_{A}\otimes\hh_{B}$ be constructed from these sets according
to Definition \ref{def1}. If state $\bro_{AB}$ is separable and
$1/\alpha+1/\beta=2$, then
\begin{align}
R_{\alpha}(\clm^{(1)}|\bro_{AB})+R_{\beta}(\clm^{(2)}|\bro_{AB})
&\geq-2\,\ln\eta_{S}
\, , \label{resep}\\
H_{\alpha}(\clm^{(1)}|\bro_{AB})+H_{\beta}(\clm^{(2)}|\bro_{AB})
&\geq\ln_{\mu}\bigl(\eta_{S}^{-2}\bigr)
\, , \label{tsasep}
\end{align}
where $S=A,B$, maximal entropic parameter
$\mu=\max\{\alpha,\beta\}$, and
$\eta_{S}=\eta(\clf_{S}^{(1)},\clf_{S}^{(2)})$ according to
(\ref{mupr1}).
\end{musep}

{\bf Proof.} We will further assume that $\alpha>1>\beta$. The
Shannon case $\alpha=\beta=1$ is finally reached by taking the
corresponding limit. It is sufficient to prove (\ref{resep}) and
(\ref{tsasep}) only for one of the cases $S=A,B$. For brevity, we
also denote
\begin{equation}
Q_{AB}^{(t)}=p(\clm^{(t)}|\bro_{AB})
\, , \qquad
p_{A}^{(t)}=p(\clf_{A}^{(t)}|\bro_{A})
\, , \qquad
q_{B}^{(t)}=p(\clf_{B}^{(t)}|\bro_{B})
\, , \label{brev}
\end{equation}
where $t=1,2$, reduced densities $\bro_{A}=\Tr_{B}(\bro_{AB})$ and
$\bro_{B}=\Tr_{A}(\bro_{AB})$.

We first suppose that $\bro_{AB}$ is a product state appeared as
$\bro_{A}\otimes\bro_{B}$. Combining
$Q_{AB}^{(t)}=p_{A}^{(t)}*q_{B}^{(t)}$ with (\ref{fgalp}) and
(\ref{fgbet}), for $\alpha>1>\beta$ we have the inequality
\begin{align}
\bigl\|Q_{AB}^{(1)}\bigr\|_{\alpha}\leq
\|p_{A}^{(1)}\|_{\alpha}\leq\eta_{A}^{2(1-\beta)/\beta}\,\|p_{A}^{(2)}\|_{\beta}
\leq\eta_{A}^{2(1-\beta)/\beta}\,\bigl\|Q_{AB}^{(2)}\bigr\|_{\beta}
\label{cmu11}
\end{align}
and its ``twin'' with swapped $Q_{AB}^{(1)}$ and $Q_{AB}^{(2)}$.
In general, we cannot assume concavity of the R\'{e}nyi
$\alpha$-entropy for $\alpha>1$. Hence, we should extend our
results to separable states before obtaining final entropic
inequalities.

Each separable state can be represented as a convex combination of
product states,
\begin{equation}
\bro_{AB}=\sum\nolimits_{\lambda} \lambda\>\bro_{A\lambda}\otimes\bro_{B\lambda}
\, . \nonumber
\end{equation}
Here, density matrices are all normalized so that
$\sum_{\lambda}\lambda=1$. For the above combination of product
states, we obtain
\begin{equation}
Q_{AB}^{(t)}=\sum\nolimits_{\lambda}\lambda\,Q_{AB\lambda}^{(t)}
\, , \label{wdenp}
\end{equation}
where each $Q_{AB\lambda}^{(t)}$ corresponds to the product
$\bro_{A\lambda}\otimes\bro_{B\lambda}$. Following
\cite{IBB06,rast10r}, at this step we use the Minkowski
inequality. Assuming $\alpha>1>\beta>0$, this inequality gives
\begin{align}
&\bigl\|Q_{AB}^{(1)}\bigr\|_{\alpha}
=\left\|\sum\nolimits_{\lambda}\lambda\,Q_{AB\lambda}^{(1)}\right\|_{\alpha}
\leq\sum\nolimits_{\lambda}\lambda\,\bigl\|Q_{AB\lambda}^{(1)}\bigr\|_{\alpha}
\, , \label{wmink}\\
&\sum\nolimits_{\lambda}\lambda\,\bigl\|Q_{AB\lambda}^{(2)}\bigr\|_{\beta}
\leq\left\|\sum\nolimits_{\lambda}\lambda\,Q_{AB\lambda}^{(2)}\right\|_{\beta}
=\bigl\|Q_{AB}^{(2)}\bigr\|_{\beta}
\, . \label{umnik}
\end{align}
For each $\lambda$, the quantities
$\bigl\|Q_{AB\lambda}^{(1)}\bigr\|_{\alpha}$ and
$\bigl\|Q_{AB\lambda}^{(2)}\bigr\|_{\beta}$ obey (\ref{cmu11}). By
(\ref{wmink}) and (\ref{umnik}), the relation (\ref{cmu11}) and
its ``twin'' written in terms of $Q_{AB}^{(t)}$ are also valid for
all separable state.

To complete the proof, we shall convert (\ref{cmu11}) into
entropic inequalities. The R\'{e}nyi entropies are represented via
norm-like functionals according to (\ref{rpdf1}). To get
(\ref{resep}) with $\alpha>1>\beta$, we take the logarithm of both
the sides of (\ref{cmu11}) and use the link
$(\alpha-1)/\alpha=(1-\beta)/\beta$. The inequality with swapped
probability distributions is then obtained by a parallel argument.
Together, these inequality are joined into (\ref{resep}) under the
condition $1/\alpha+1/\beta=2$ solely. The case of Tsallis
entropies is not so immediate. Following \cite{rast104}, we can
examine a minimization problem under the restrictions imposed by
(\ref{cmu11}) and its ``twin''. Calculations resulting in
(\ref{tsasep}) are very similar to the derivation given in
appendix of \cite{rast104}. $\blacksquare$

The statement of Proposition \ref{pon41} provides entropic
separability conditions based on local uncertainty relations of
the Maassen--Uffink type. These formulas hold under the
restriction $1/\alpha+1/\beta=2$. The latter reflects the fact
that the Maassen--Uffink result is derived from Riesz's theorem
\cite{riesz27}. An alternative viewpoint is that such uncertainty
relations follow from the monotonicity of the quantum relative
entropy \cite{ccyz12}. The used scheme of building total
measurements also leads to separability conditions on the base
of majorization uncertainty relations. Our second result is
posed as follows.

\newtheorem{majsep}[fojen]{Proposition}
\begin{majsep}\label{pon42}
Let each of sets $\clf_{A}^{(1)}$ and $\clf_{A}^{(2)}$ of
subnormalized vectors form rank-one POVM in $\hh_{A}$, and let
each of sets $\clf_{B}^{(1)}$ and $\clf_{B}^{(2)}$ of
subnormalized vectors form rank-one POVM in $\hh_{B}$. Let two
POVMs $\clm^{(t)}(\clf_{A}^{(t)},\clf_{B}^{(t)})$ in
$\hh_{A}\otimes\hh_{B}$ be constructed from these sets according
to Definition \ref{def1}. For $S=A,B$, we introduce $D\times{D}$
matrix $\vm_{S}=\vm(\clf_{S}^{(1)},\clf_{S}^{(2)})$ due to
(\ref{neww}). To each of such two matrices, we assign the sequence
of numbers according to (\ref{skdf}) and the majorizing vector
$w_{S}$, where $S=A,B$. For each separable state $\bro_{AB}$ and
$0<\alpha\leq1$, there holds
\begin{equation}
R_{\alpha}(\clm^{(1)}|\bro_{AB})+R_{\alpha}(\clm^{(2)}|\bro_{AB})
\geq{R}_{\alpha}(w_{S})
\, . \label{remaj}
\end{equation}
For each separable state $\bro_{AB}$ and $\alpha>0$, there holds
\begin{equation}
H_{\alpha}(\clm^{(1)}|\bro_{AB})+H_{\alpha}(\clm^{(2)}|\bro_{AB})
\geq{H}_{\alpha}(w_{S})
\, . \label{tsamaj}
\end{equation}
\end{majsep}

{\bf Proof.} In view of symmetry between subsystems, we will prove
(\ref{remaj}) and (\ref{tsamaj}) only for one of the cases
$S=A,B$. Combining (\ref{pqmaq}) with (\ref{pppcn1}) and using the
notation (\ref{brev}) again, for each product state we write
\begin{align}
R_{\alpha}\bigl(Q_{AB}^{(t)}\bigr)&\geq{R}_{\alpha}\bigl(p_{A}^{(t)}\bigr)
\, , \nonumber\\
H_{\alpha}\bigl(Q_{AB}^{(t)}\bigr)&\geq{H}_{\alpha}\bigl(p_{A}^{(t)}\bigr)
\, . \nonumber
\end{align}
It is essential here that both the $\alpha$-entropies are
Schur-concave. By the majorization-based relation
(\ref{nwmr0}) and (\ref{nwmr01}), for a product state we have
\begin{align}
R_{\alpha}\bigl(Q_{AB}^{(1)}\bigr)+R_{\alpha}\bigl(Q_{AB}^{(2)}\bigr)\geq
{R}_{\alpha}\bigl(p_{A}^{(1)}\bigr)+R_{\alpha}\bigl(p_{A}^{(2)}\bigr)&\geq{R}_{\alpha}(w_{A})
&(0<\alpha\leq1)
\, , \label{mars2}\\
H_{\alpha}\bigl(Q_{AB}^{(1)}\bigr)+H_{\alpha}\bigl(Q_{AB}^{(2)}\bigr)\geq
{H}_{\alpha}\bigl(p_{A}^{(1)}\bigr)+H_{\alpha}\bigl(p_{A}^{(2)}\bigr)&\geq{H}_{\alpha}(w_{A})
&(0<\alpha<\infty)
\, . \label{mats2}
\end{align}
As the R\'{e}nyi $\alpha$-entropy is concave for $0<\alpha\leq1$,
the formula (\ref{mars2}) implies (\ref{remaj}) for all separable
states. The claim (\ref{tsamaj}) follows from (\ref{mats2}) due to
concavity of the Tsallis $\alpha$-entropy. $\blacksquare$

In Proposition \ref{pon42}, we deal with separability conditions
derived from majorization uncertainty relations. Unlike
(\ref{resep}), the condition (\ref{remaj}) is restricted to the
range $0<\alpha\leq1$, where R\'{e}nyi's entropy is concave
irrespectively to dimensionality of probabilistic vectors. For
$\alpha>1$, concavity properties actually depend on dimensionality
of probabilistic vectors. So, the binary R\'{e}nyi entropy is
concave for all $0<\alpha\leq2$ \cite{ben78}. Since the case of
qubits is very important, we give two separability conditions
additional to (\ref{remaj}). Here, the majorization-based
relations (\ref{nwmr1}) and (\ref{oldmr}) will be used.

Assuming $D=2$, we consider pairs
$\bigl\{\cle_{A}^{(1)},\cle_{A}^{(2)}\bigr\}$ and
$\bigl\{\cle_{B}^{(1)},\cle_{B}^{(2)}\bigr\}$ of orthonormal
bases, each in two dimensions. For product states of a two-qubit
system and $\alpha>1$, one has
\begin{align}
R_{\alpha}\bigl(Q_{AB}^{(1)}\bigr)+R_{\alpha}\bigl(Q_{AB}^{(2)}\bigr)
&\geq\frac{2}{1-\alpha}{\>}
\ln\!\left(
\frac{1+\|w_{S}\|_{\alpha}^{\alpha}}{2}
\right)
 , \label{mars22}\\
R_{\alpha}\bigl(Q_{AB}^{(1)}\bigr)+R_{\alpha}\bigl(Q_{AB}^{(2)}\bigr)
&\geq{R}_{\alpha}(w_{S}^{\prime})
\, . \label{mars23}
\end{align}
The vectors $\omega_{S}$ and $\omega_{S}^{\prime}$ are
obtained for unitary $2\times2$ matrix
$\vm_{S}=\vm(\cle_{S}^{(1)},\cle_{S}^{(2)})$ with $S=A,B$ in line
with (\ref{wdd1}) and (\ref{wpdd}). We cannot extend
(\ref{mars22}) and (\ref{mars23}) to separable states without
entropic concavity. When $d=2$ and $1<\alpha\leq2$, for each
separable state $\bro_{AB}$ we finally get
\begin{align}
R_{\alpha}(\clm^{(1)}|\bro_{AB})+R_{\alpha}(\clm^{(2)}|\bro_{AB})
&\geq\frac{2}{1-\alpha}{\>}
\ln\!\left(
\frac{1+\|w_{S}\|_{\alpha}^{\alpha}}{2}
\right)
 , \label{smar22}\\
R_{\alpha}(\clm^{(1)}|\bro_{AB})+R_{\alpha}(\clm^{(2)}|\bro_{AB})
&\geq{R}_{\alpha}(w_{S}^{\prime})
\, . \label{smar23}
\end{align}
For a two-qubit system, majorization-based separability conditions
in terms of R\'{e}nyi's entropies are given in the range
$0<\alpha\leq2$.

We shall now proceed to separability conditions related to
local measurements with a special structure. In the case of MUBs,
the following statement takes place.

\newtheorem{mamub}[fojen]{Proposition}
\begin{mamub}\label{pon43}
Let $\bigl\{\cle_{A}^{(1)},\ldots,\cle_{A}^{(K)}\bigr\}$ be a set
of $K$ MUBs in $\hh_{A}$. Let
$\bigl\{\cle_{B}^{(1)},\ldots,\cle_{B}^{(K)}\bigr\}$ be a set of
$K$ MUBs in $\hh_{B}$. Let $K$ POVMs
$\clm^{(t)}(\cle_{A}^{(t)},\cle_{B}^{(t)})$ in
$\hh_{A}\otimes\hh_{B}$ be constructed from these MUBs according
to Definition \ref{def1}. For each separable state $\bro_{AB}$ and
$\alpha\in(0;2]$, there holds
\begin{equation}
\frac{1}{K}\,\sum_{t=1}^{K} H_{\alpha}(\clm^{(t)}|\bro_{AB})
\geq\ln_{\alpha}\!\left(\frac{Kd}{d+K-1}\right)
 , \label{scmub02}
\end{equation}
where $d=\dmn(\hh_{A})=\dmn(\hh_{B})$.
\end{mamub}

{\bf Proof.} Again, we will prove (\ref{scmub02}) only for one of
the cases $S=A,B$. The Tsallis $\alpha$-entropy is Schur-concave
for all $\alpha\in(0;2]$. Combining this fact with (\ref{pqmaq})
and (\ref{tmub02}), for any product we have
\begin{equation}
\frac{1}{K}\,\sum_{t=1}^{K} H_{\alpha}(\clm^{(t)}|\bro_{A}\otimes\bro_{B})
\geq\frac{1}{K}\,\sum_{t=1}^{K} H_{\alpha}(\cle_{A}^{(t)}|\bro_{A})
\geq\ln_{\alpha}\!\left(\frac{Kd}{d+K-1}\right)
 . \label{tubab}
\end{equation}
By concavity of Tsallis' $\alpha$-entropy, we extend the latter to
all separable states. $\blacksquare$

In the particular case $\alpha=1$, we have
\begin{equation}
\frac{1}{K}\,\sum_{t=1}^{K} H_{1}(\clm^{(t)}|\bro_{AB})
\geq\ln\!\left(\frac{Kd}{d+K-1}\right)
 , \nonumber
\end{equation}
whenever $\bro_{AB}$ is separable. This separability condition is
actually those that can be derived from the R\'{e}nyi-entropy
bound (\ref{rmub02}). When $d$ is not specified, we can use
concavity of R\'{e}nyi's $\alpha$-entropy only for
$0<\alpha\leq1$. In addition, it does not increase with growth of
$\alpha$. For $d=2$, however, the R\'{e}nyi $\alpha$-entropy is
concave up to $\alpha=2$. Thus, for a two-qubit system we write
the condition
\begin{equation}
\frac{1}{K}\,\sum_{t=1}^{K} R_{2}(\clm^{(t)}|\bro_{AB})
\geq\ln\!\left(\frac{2K}{K+1}\right)
 , \label{tubab2}
\end{equation}
where $K=2,3$ and $\bro_{AB}$ is separable. The result remains
formally valid for $K=1$, but the bound becomes trivial here.

Detecting entanglement, we should use as many complementary
measurement as possible. When the dimensionality $d$ of subsystems
is a prime power, $d+1$ mutually unbiased bases exist. For other
values of $d$, we may apply mutually unbiased measurements, since
a complete set of $d+1$ MUMs exists in all dimensions
\cite{kagour}. Hence, entanglement detection with MUMs is of
interest. The following claim is derived from (\ref{tmum02})
similarly to the proof of Proposition \ref{pon43}.

\newtheorem{mamum}[fojen]{Proposition}
\begin{mamum}\label{pon44}
Let $\bigl\{\cln_{A}^{(1)},\ldots,\cln_{A}^{(K)}\bigr\}$ be a set
of $K$ MUMs of the efficiency $\varkappa_{A}$ in $\hh_{A}$, and
let $\bigl\{\cln_{B}^{(1)},\ldots,\cln_{B}^{(K)}\bigr\}$ be a set
of $K$ MUMs of the efficiency $\varkappa_{B}$ in $\hh_{B}$. Let
$K$ POVMs $\clm^{(t)}(\cln_{A}^{(t)},\cln_{B}^{(t)})$ in
$\hh_{A}\otimes\hh_{B}$ be constructed from these MUMs according
to Definition \ref{def1}. For each separable state $\bro_{AB}$ and
$\alpha\in(0;2]$, there holds
\begin{equation}
\frac{1}{K}\,\sum_{t=1}^{K} H_{\alpha}(\clm^{(t)}|\bro_{AB})
\geq\ln_{\alpha}\!\left(\frac{Kd}{\varkappa_{S}d+K-1}\right)
 , \label{scmum02}
\end{equation}
where $S=A,B$ and $d=\dmn(\hh_{A})=\dmn(\hh_{B})$.
\end{mamum}

Entropic uncertainty relations for symmetric informationally
complete measurements also lead to separability conditions. Note
that such separability conditions are formulated for a single
measurement. We give formulations for a usual SIC-POVM and then
for a general one.

\newtheorem{masic}[fojen]{Proposition}
\begin{masic}\label{pon45}
Let $\clf_{A}$ and $\clf_{B}$ be two sets of subnormalized vectors
that form SIC-POVMs in $\hh_{A}$ and $\hh_{B}$, respectively. Let
POVM $\clm(\clf_{A},\clf_{B})$ in $\hh_{A}\otimes\hh_{B}$ be
constructed according to Definition \ref{def1}. For each separable
state $\bro_{AB}$ and $\alpha\in(0;2]$, there holds
\begin{equation}
H_{\alpha}(\clm|\bro_{AB})
\geq\ln_{\alpha}\!\left(\frac{d(d+1)}{2}\right)
 , \label{scsic02}
\end{equation}
where $d=\dmn(\hh_{A})=\dmn(\hh_{B})$.
\end{masic}

{\bf Proof.} The Tsallis $\alpha$-entropy is Schur-concave for all
$\alpha\in(0;2]$. Combining this fact with (\ref{pqmaq}) and
(\ref{tsic02}), for any product we have
\begin{equation}
H_{\alpha}(\clm|\bro_{A}\otimes\bro_{B})
\geq{H}_{\alpha}(\clf_{A}|\bro_{A})
\geq\ln_{\alpha}\!\left(\frac{d(d+1)}{2}\right)
 . \label{tsiab}
\end{equation}
By concavity of Tsallis' $\alpha$-entropy, the latter is extended
to all separable states. $\blacksquare$

General SIC-POVM exist in all finite dimensions \cite{kgour13}.
Moreover, they can be obtained within a unifying framework. Even
if usual SIC-POVMs exist in all dimensions, they may be difficult
to implement. Thus, entanglement detection with general SIC-POVM
may be more appropriate. The following claim is derived from
(\ref{tgsic02}) similarly to the proof of Proposition \ref{pon45}.

\newtheorem{magsic}[fojen]{Proposition}
\begin{magsic}\label{pon46}
Let $\cln_{A}$ and $\cln_{B}$ be two general SIC-POVMs in
$\hh_{A}$ and $\hh_{B}$, respectively. Let POVM
$\clm(\cln_{A},\cln_{B})$ in $\hh_{A}\otimes\hh_{B}$ be
constructed according to Definition \ref{def1}. For each separable
state $\bro_{AB}$ and $\alpha\in(0;2]$, there holds
\begin{equation}
H_{\alpha}(\clm|\bro_{AB})
\geq\ln_{\alpha}\!\left(\frac{d(d+1)}{a_{S}d^{2}+1}\right)
 , \label{scgic02}
\end{equation}
where $S=A,B$ and $d=\dmn(\hh_{A})=\dmn(\hh_{B})$.
\end{magsic}

We have derived a lot of separability conditions in terms of the
R\'{e}nyi and Tsallis entropies. The violation of any separability
condition certifies that the measured state of a bipartite system
is entangled. We also note that the presented conditions actually
concern biseparability. The problem of detecting multipartite
entanglement is generally more complicated
\cite{popescu98,ibkz16}. This issue was addressed in
\cite{toth06,huangy12,huber13,zhao15}.

\section{Discussion}\label{sec5}

In this section, we will apply the presented conditions to some
states, for which separability limits are already known. We also
compare the two types of derived criteria with each other and also
with previous criteria given in the literature. Bipartite
separability conditions are often tested with density matrices of
the form
\begin{equation}
(1-c)\vbro_{sep}+c\,|\Phi\rangle\langle\Phi|
\, . \nonumber
\end{equation}
Here, the density matrix $\vbro_{sep}$ is separable,
$|\Phi\rangle$ is a maximally entangled state, and real constant
$c\in[0;1]$. Taking $\vbro_{sep}$ to be the completely mixed
state, the above form is a bipartite case of Werner states
\cite{werner89}. A bipartite Werner state is separable if and only
if \cite{rubin2000}
\begin{equation}
c\leq\frac{1}{d+1}
\ , \label{crub}
\end{equation}
where $d$ is the dimensionality of each of two subsystems.
The authors of \cite{rubin2000} also presented necessary and
sufficient conditions for multipartite Werner states.

We begin with a two-qubit system. The corresponding Pauli matrices
are denoted as $\bsg_{x}$, $\bsg_{y}$, $\bsg_{z}$. For a bipartite
system of two qubits, the inequality (\ref{crub}) gives
$c\leq1/3$. The entangled pure state $|\Phi\rangle$ will be taken
as
\begin{equation}
|\Phi\rangle=\frac{1}{\sqrt{2}}
\,\bigl(
|z_{0}z_{0}\rangle+|z_{1}z_{1}\rangle
\bigr)
\, , \label{psid}
\end{equation}
where $\bigl\{|z_{0}\rangle,|z_{1}\rangle\bigr\}$ is the
eigenbasis of $\bsg_{z}$. We further consider a family of density
matrices
\begin{equation}
\vbro_{\Phi}=\frac{1-c}{4}\>\pen_{2}\otimes\pen_{2}+c\,|\Phi\rangle\langle\Phi|
\, . \label{werrd}
\end{equation}
Let us take the basis $\bigl\{|z_{0}\rangle,|z_{1}\rangle\bigr\}$  and a
rotated basis $\bigl\{|u_{0}\rangle,|u_{1}\rangle\bigr\}$ such
that
\begin{equation}
|u_{0}\rangle:=\cos\theta\,|z_{0}\rangle+\sin\theta\,|z_{1}\rangle
\, , \qquad
|u_{1}\rangle:=\sin\theta\,|z_{0}\rangle-\cos\theta\,|z_{1}\rangle
\, , \label{u01df}
\end{equation}
where $\theta\neq0$. So, we wish to deal not only with two
mutually unbiased bases. This point is essential for comparing
majorization-based separability conditions with separability
conditions of the Maassen--Uffink type. For each of two qubits,
we actually take the observables $\bsg_{z}$ and
\begin{equation}
|u_{0}\rangle\langle{u}_{0}|-|u_{1}\rangle\langle{u}_{1}|
\, . \label{ambn2}
\end{equation}
For $\theta=\pi/4$, the second basis
$\bigl\{|u_{0}\rangle,|u_{1}\rangle\bigr\}$ gives the eigenbasis
of $\bsg_{x}$, whence (\ref{ambn2}) reads as it. The
total measurements $\clm^{(z)}$ and $\clm^{(u)}$ respectively contain
the projectors
\begin{align}
&\lasf_{0}^{(z)}=|z_{0}z_{0}\rangle\langle{z}_{0}z_{0}|+
|z_{1}z_{1}\rangle\langle{z}_{1}z_{1}|
\, ,
&\lasf_{1}^{(z)}=|z_{0}z_{1}\rangle\langle{z}_{0}z_{1}|+
|z_{1}z_{0}\rangle\langle{z}_{1}z_{0}|
\, , \label{laf11}\\
&\lasf_{0}^{(u)}=|u_{0}u_{0}\rangle\langle{u}_{0}u_{0}|+
|u_{1}u_{1}\rangle\langle{u}_{1}u_{1}|
\, ,
&\lasf_{1}^{(u)}=|u_{0}u_{1}\rangle\langle{u}_{0}u_{1}|+
|u_{1}u_{0}\rangle\langle{u}_{1}u_{0}|
\, . \label{laf22}
\end{align}
For $\theta$ in the quadrant I, we have
$\eta_{S}=\max\{\cos\theta,\sin\theta\}\equiv\eta$ and
\begin{align}
w&=\bigl(\eta,1-\eta\bigr)
\, , \label{omebn}\\
w^{\prime}&=\frac{1}{4}\>\bigl(1+2\eta+\eta^{2},3-2\eta-\eta^{2}\bigr)
\, . \label{omepbn}
\end{align}
Combining these observations with (\ref{resep}), (\ref{tsasep}),
(\ref{remaj}), (\ref{tsamaj}), (\ref{smar22}), and (\ref{smar23}),
we obtain a lot of separability conditions for the case
considered.

When $\theta=\pi/4$, we have $\eta=1/\sqrt{2}$ and two MUBs,
namely the eigenbases of $\bsg_{z}$ and $\bsg_{x}$.  By
calculations, we obtain
\begin{equation}
\langle\Phi|\lasf_{i}^{(z)}|\Phi\rangle=\delta_{i0}
\, , \qquad
\langle\Phi|\lasf_{j}^{(x)}|\Phi\rangle=\delta_{j0}
\, , \nonumber
\end{equation}
where $\delta_{ij}$ is the Kronecker symbol. On the completely
mixed state, each of two measurements generates the uniform
distribution. For the state (\ref{werrd}), we twice obtain the
pair of probabilities $(1\pm{c})/2$. By inspection, the best
detection among relations of the Maassen--Uffink type is provided
by (\ref{resep}) for the choice $\alpha=\infty$ and $\beta=1/2$.
Here, we have the condition
\begin{equation}
-\ln\!\left(\frac{1+c}{2}\right)+\ln\!\left(1+\sqrt{1-c^{2}}\right)
\geq\ln2
\, , \nonumber
\end{equation}
which is equivalent to $c\leq1/\sqrt{2}$. So, the entropic
separability conditions of the form (\ref{resep}) detect
entanglement when $c>1/\sqrt{2}\approx0.7071$. The same range
takes place for the criterion that considers the sum of maximal
probabilities in two measurements. It is rather natural since that
criterion is also based on the Maassen--Uffink approach
\cite{devic05}. The result quoted follows from a general
formulation by substituting $d=2$. Performing a direct
optimization in the qubit case allows us to improve restrictions
\cite{devic05}. On the other hand, this approach becomes hardly
appropriate with growth of the dimensionality.

Let us proceed to majorization-based separability conditions. The
condition (\ref{smar22}) gives
\begin{equation}
-\ln\!\left(\frac{1+c^{2}}{2}\right)\geq
-\ln\!\left(\frac{1+\|w\|_{2}^{2}}{2}\right)
 , \label{renaj2}
\end{equation}
where $\|w\|_{2}^{2}=2-\sqrt{2}$. With (\ref{renaj2}), we are able
to detect entanglement for $c>\|w\|_{2}\approx0.7654$. Calculating
$\|w^{\prime}\|_{2}$ for $\eta=1/\sqrt{2}$, the condition
(\ref{smar23}) reads
\begin{equation}
-\ln\!\left(\frac{1+c^{2}}{2}\right)\geq
-\ln\|w^{\prime}\|_{2}
\, . \label{renaj3}
\end{equation}
For $c>\sqrt{2\,\|w^{\prime}\|_{2}-1}\approx0.7450$, we can detect
entanglement due to (\ref{renaj3}). For both the
majorization-based conditions, the range of detection is slightly
less than for separability conditions of the Maassen--Uffink type.

It is instructive to address bases that are not mutually unbiased.
With $\theta=\pi/6$, the best result among conditions of the
Maassen--Uffink type is reached by (\ref{resep}) for the choice
$\alpha=\beta=1$. Separability conditions of the form
(\ref{resep}) detect entanglement for $c>c_{1}$ with
$c_{1}\approx0.9347$. The best result among majorization-based
conditions is provided due to (\ref{renaj3}). With the latter, we
are able to detect entanglement for $c>c_{2}$ with
$c_{2}\approx0.8719$. Using the majorization-based separability
conditions, we see more effective detection. For other $\theta$,
results were found to be similar. Separability conditions of the
Maassen--Uffink type are rather preferable for MUBs. Of course,
such bases are used in many schemes for entanglement detection
{\sl per se}. However, in quantum information science we may also
perform measurements designed for other purposes. Statistics of
such measurements may nevertheless be used additionally for
entanglement detection. Here, we can apply separability conditions
based on majorization uncertainty relations.

Let us compare two forms of separability conditions with using
three MUBs. For a qubit, these MUBs are taken as the eigenbases of
the Pauli observables. The first form deals with the so-called
correlation measure introduced in \cite{shbah12}. Using
(\ref{indk}), the authors of \cite{shbah12} obtained separability
conditions in terms of the correlation measure. In two dimensions,
the correlation measure is expressed as
\begin{equation}
J(\bro_{AB})=\sum_{t=z,x,y}\,\sum_{i=0,1}
\langle{e}_{i}^{(t)}e_{i}^{(t)}|\bro_{AB}|e_{i}^{(t)}e_{i}^{(t)}\rangle
\, , \label{cojdf}
\end{equation}
where
$|ee^{\prime}\rangle\equiv|e\rangle\otimes|e^{\prime}\rangle$. For
separable states of a two-qubit system, the correlation measure
satisfies $J\leq2$. Simple calculations now give
\begin{equation}
J(\vbro_{\Phi})=\frac{1+c}{2}+\frac{1+c}{2}+\frac{1-c}{2}=\frac{3+c}{2}
\ . \label{cormub}
\end{equation}
For all $c\in[0;1]$, the right-hand side of (\ref{cormub}) does
not violate the separability condition $J\leq2$. Density matrices
of the form (\ref{werrd}) are not separable for $c>1/3$ and all
escape the entanglement detection with respect to this criterion.
It is not the case for separability conditions such as
(\ref{scmub02}) and (\ref{tubab2}).

For each of three measurements
$\clm^{(t)}=\bigl\{\lasf_{0}^{(t)},\lasf_{1}^{(t)}\bigr\}$, where
$t=z,x,y$, the projectors are written according to (\ref{laf11}).
By direct calculations, we have
\begin{equation}
\langle\Phi|\lasf_{0}^{(z)}|\Phi\rangle=1
\, , \qquad
\langle\Phi|\lasf_{0}^{(x)}|\Phi\rangle=1
\, , \qquad
\langle\Phi|\lasf_{1}^{(y)}|\Phi\rangle=1
\, . \label{lasfzxy}
\end{equation}
On the completely mixed state, each of measurements generates the
uniform distribution with two outcomes. For the state
(\ref{werrd}), we obtain probabilities $(1\pm{c})/2$ in all three
cases. Substituting $d=2$, $K=3$ and $\alpha=2$, both the entropic
bounds (\ref{scmub02}) and (\ref{tubab2}) lead to the condition
\begin{equation}
\left(\frac{1+c}{2}\right)^{\!2}+\left(\frac{1-c}{2}\right)^{\!2}
=\frac{1+c^{2}}{2}\leq\frac{2}{3}
\ , \nonumber
\end{equation}
or merely $c\leq1/\sqrt{3}$. So, the entropic separability
conditions (\ref{scmub02}) and (\ref{tubab2}) detect entanglement
for $c>1/\sqrt{3}\approx0.5774$. In the considered example, the
entropic approach is more effective than the method based on
the correlation measure. An efficiency of separability conditions
is very sensitive to the choice of local measurement bases. For
conditions in terms of maximal probabilities, this fact was
already mentioned in \cite{rastsep}. Further, the range
$c>1/\sqrt{3}$ is wider than the range $c>1/\sqrt{2}$, in which
separability conditions of the form (\ref{resep}) are able to
detect entanglement. Our abilities to detect entanglement should
increase, when the number of involved bases grows and used
separability conditions are chosen properly.

Using several MUBs, we have two possible types of separability
conditions, one in terms of entropies and another in terms of
correlation measures. It is also instructive to compare these
types with entangled states of a two-qutrit system. Let us recall
the generalized Pauli operators
\begin{equation}
\az=\begin{pmatrix}
1 & 0 & 0 \\
0 & \,\omega_{3} & 0 \\
0 & 0 & \omega_{3}^{*}
\end{pmatrix}
 ,
\qquad
\ax=\begin{pmatrix}
0\, & 0 & 1 \\
1 & 0\, & 0 \\
0 & 1 & 0\,
\end{pmatrix}
 , \nonumber
\end{equation}
where $\omega_{3}=\exp(\iu2\pi/3)$. Four MUBs in the qutrit
Hilbert space can be described as the eigenbases of the operators
$\az$, $\ax$, $\az\ax$, and $\az\ax^{2}$. We shall test
separability conditions on the following family of states,
\begin{align}
\vbro_{\Psi}&=\frac{1-c}{9}\>\pen_{3}\otimes\pen_{3}+c\,|\Psi\rangle\langle\Psi|
\, , \label{dwerr}\\
|\Psi\rangle&=\frac{1}{\sqrt{3}}
\,\bigl(
|z_{0}x_{0}\rangle+|z_{1}x_{2}\rangle+|z_{2}x_{1}\rangle
\bigr)
\, . \label{dpsi}
\end{align}
By $\bigl\{|z_{i}\rangle\bigr\}$, $\bigl\{|x_{i}\rangle\bigr\}$,
$\bigl\{|y_{i}\rangle\bigr\}$, with $i=0,1,2$, we respectively
mean the eigenbases of $\az$, $\ax$, and $\az\ax$. Defining the
correlation measure for three MUBs similarly to (\ref{cojdf}),
one obtains $J(\vbro_{\Psi})=1$. Separability
conditions in terms of correlation measures follow from
(\ref{indk}). When $d=3$ and $K=3$, for all separable states we
have the condition $J\leq5/3$. The latter is fulfilled by
$J(\vbro_{\Psi})$ independently of $c\in[0;1]$. Density matrices
of the form (\ref{dwerr}) are not separable for $c>1/4$ and all
escape the entanglement detection with respect to this criterion.
Note that $|\Psi\rangle$ is an eigenstate of three operators,
so that
\begin{equation}
(\az\otimes\ax)|\Psi\rangle=|\Psi\rangle
\, , \qquad
(\ax\otimes\az)|\Psi\rangle=|\Psi\rangle
\, , \qquad
(\az\ax\otimes\az\ax)|\Psi\rangle=\omega_{3}\,|\Psi\rangle
\, . \label{3eigv}
\end{equation}
Hence, we may try to rotate unitarily local measurement bases. For
instance, one can take simultaneously $\bigl\{|z_{i}\rangle\bigr\}$ on
qutrit $A$ and $\bigl\{|x_{i}\rangle\bigr\}$ on qutrit $B$, and so on. 
Calculating the measure
\begin{equation}
\sum_{i=0,1,2}
\Bigl(
\langle{z}_{i}x_{i}|\vbro_{\Psi}|z_{i}x_{i}\rangle+
\langle{x}_{i}z_{i}|\vbro_{\Psi}|x_{i}z_{i}\rangle+
\langle{y}_{i}y_{i}|\vbro_{\Psi}|y_{i}y_{i}\rangle
\Bigr)=1
\, , \nonumber
\end{equation}
we still see no violation.

Let us consider three measurements designed according to
Definition \ref{def1}. To the first pair of bases, we assign the
three projectors
\begin{align}
\pisf_{0}^{(zx)}&=|z_{0}x_{0}\rangle\langle{z}_{0}x_{0}|+
|z_{1}x_{2}\rangle\langle{z}_{1}x_{2}|+|z_{2}x_{1}\rangle\langle{z}_{2}x_{1}|
\, , \nonumber\\
\pisf_{1}^{(zx)}&=|z_{0}x_{1}\rangle\langle{z}_{0}x_{1}|+
|z_{1}x_{0}\rangle\langle{z}_{1}x_{0}|+|z_{2}x_{2}\rangle\langle{z}_{2}x_{2}|
\, , \nonumber\\
\pisf_{2}^{(zx)}&=|z_{0}x_{2}\rangle\langle{z}_{0}x_{2}|+
|z_{1}x_{1}\rangle\langle{z}_{1}x_{1}|+|z_{2}x_{0}\rangle\langle{z}_{2}x_{0}|
\, , \nonumber
\end{align}
which form $\clm^{(zx)}$. In a similar manner, we write projectors
of the measurements $\clm^{(xz)}=\bigl\{\pisf_{k}^{(xz)}\bigr\}$
and $\clm^{(yy)}=\bigl\{\pisf_{k}^{(yy)}\bigr\}$. It immediately
follows from (\ref{3eigv}) that
\begin{equation}
\langle\Psi|\pisf_{0}^{(zx)}|\Psi\rangle=1
\, , \qquad
\langle\Psi|\pisf_{0}^{(xz)}|\Psi\rangle=1
\, , \qquad
\langle\Psi|\pisf_{1}^{(yy)}|\Psi\rangle=1
\, . \nonumber
\end{equation}
On the completely mixed state, each of three measurements
generates the uniform distribution with three outcomes. For the
state (\ref{dwerr}), we have three probability distributions,
each with one entry $(1+2c)/3$ and two entries $(1-c)/3$.
Substituting $d=3$, $K=3$ and $\alpha=2$, the entropic bound
(\ref{scmub02}) gives the condition
\begin{equation}
\left(\frac{1+2c}{3}\right)^{\!2}+2\left(\frac{1-c}{3}\right)^{\!2}
=\frac{1+2c^{2}}{3}\leq\frac{5}{9}
\ , \nonumber
\end{equation}
or $c\leq1/\sqrt{3}$. So, the entropic separability condition
(\ref{scmub02}) can detect entanglement of (\ref{dwerr}) when
$c>1/\sqrt{3}\approx0.5774$. We again observe cases, in which the
derived separability conditions are more efficient than
separability conditions in terms of correlation measures. Of
course, both the types essentially depend on local unitary
rotations and permutations of kets in bases. However, such
operations will considerably increase costs of entanglement
detection. In practice, when resources are fixed, we should
therefore try to use as many separability conditions as possible.


\begin{thebibliography}{100}

\bibitem{cat35}
Schr\"{o}dinger, E.: Die gegenw\"{a}rtige situation in der
quantenmechanik. Naturwissenschaften {\bf 23}, 807--812, 823--828,
844--849 (1935)

\bibitem{epr35}
Einstein, A., Podolsky, B., Rosen, N.: Can quantum-mechanical description of physical reality be considered
complete? Phys. Rev. {\bf 47}, 777--780 (1935)

\bibitem{hhhh09}
Horodecki, R., Horodecki, P., Horodecki, M., Horodecki, K.: Quantum entanglement. Rev. Mod. Phys. {\bf 81}, 865--942 (2009)

\bibitem{peres96}
Peres, A.: Separability criterion for density matrices. Phys. Rev. Lett. {\bf 77}, 1413--1415 (1996)

\bibitem{mhph1999}
Horodecki, M., Horodecki, P.: Reduction criterion of separability
and limits for a class of distillation protocols. Phys. Rev. A
{\bf 59}, 4206--4216 (1999)

\bibitem{horodecki96}
Horodecki, M., Horodecki, P., Horodecki, R.: Separability of mixed
states: necessary and sufficient conditions. Phys. Lett. A {\bf
223}, 1--8 (1996)

\bibitem{guhne04}
G\"{u}hne, O., Lewenstein, M.: Entropic uncertainty relations and entanglement. Phys. Rev. A {\bf 70}, 022316 (2004)

\bibitem{giovan2004}
Giovannetti, V.: Separability conditions from entropic uncertainty relations. Phys. Rev. A {\bf 70}, 012102 (2004)

\bibitem{devic05}
de Vicente, J.I., S\'{a}nchez-Ruiz, J.: Separability conditions
from the Landau--Pollak uncertainty relation. Phys. Rev. A {\bf
71}, 052325 (2005)

\bibitem{guhne06}
G\"{u}hne, O., Mechler, M., T\'{o}th, G., Adam, P.: Entanglement
criteria based on local uncertainty relations are strictly
stronger than the computable cross norm criterion. Phys. Rev. A
{\bf 74}, 010301(R) (2006)

\bibitem{devic07}
de Vicente, J.I.: Lower bounds on concurrence and separability conditions. Phys. Rev. A {\bf 75}, 052320 (2007)

\bibitem{huang2010}
Huang, Y.: Entanglement criteria via concave-function uncertainty relations. Phys. Rev. A {\bf 82}, 012335 (2010)

\bibitem{rastsep}
Rastegin, A.E.: Separability conditions based on local fine-grained uncertainty relations. Quantum Inf. Process. {\bf 15}, 2621--2638 (2016)

\bibitem{huang2013}
Huang, Y.: Entanglement detection: complexity and Shannon entropic criteria. IEEE Trans. Inf. Theor. {\bf 59}, 6774--6778 (2013)

\bibitem{heisenberg}
Heisenberg, W.: \"{U}ber den anschaulichen inhalt der quanten
theoretischen kinematik und mechanik. Zeitschrift f\"{u}r Physik
{\bf 43}, 172--198 (1927)

\bibitem{hall99}
Hall, M.J.W.: Universal geometric approach to uncertainty, entropy, and information. Phys. Rev. A {\bf 59}, 2602--2615 (1999)

\bibitem{lahti}
Busch, P., Heinonen, T., Lahti, P.J.: Heisenberg's uncertainty principle. Phys. Rep. {\bf 452}, 155--176 (2007)

\bibitem{deutsch}
Deutsch, D.: Uncertainty in quantum measurements. Phys. Rev. Lett. {\bf 50}, 631--633 (1983)

\bibitem{kraus}
Kraus, K.: Complementary observables and uncertainty relations. Phys. Rev. D {\bf 35}, 3070--3075 (1987)

\bibitem{maass}
Maassen, H., Uffink, J.B.M.: Generalized entropic uncertainty relations. Phys. Rev. Lett. {\bf 60}, 1103--1106 (1988)

\bibitem{hirs}
Hirschman, I.I.: A note on entropy. Am. J. Math. {\bf 79}, 152--156 (1957)

\bibitem{beckner}
Beckner, W.: Inequalities in Fourier analysis. Ann. Math. {\bf 102}, 159--182 (1975)

\bibitem{birula1}
Bia{\l}ynicki-Birula, I., Mycielski, J.: Uncertainty relations for
information entropy in wave mechanics. Commun. Math. Phys. {\bf
44}, 129--132 (1975)

\bibitem{ww10}
Wehner, S., Winter, A.: Entropic uncertainty relations -- a survey. New J. Phys. {\bf 12}, 025009 (2010)

\bibitem{brud11}
Bia{\l}ynicki-Birula, I., Rudnicki, {\L}.: Entropic uncertainty
relations in quantum physics. In: Sen, K.D. (ed.): Statistical
Complexity, 1--34. Springer, Berlin (2011)

\bibitem{cbtw15}
Coles, P.J., Berta, M., Tomamichel, M., Wehner, S.: Entropic
uncertainty relations and their applications. Rev. Mod. Phys. {\bf 89}, 015002 (2017)

\bibitem{oppwn10}
Oppenheim, J., Wehner, S.: The uncertainty principle determines
the nonlocality of quantum mechanics. Science {\bf 330},
1072--1074 (2010)

\bibitem{renf13}
Ren, L.-H., Fan, H.: General fine-grained uncertainty relation and the second law of thermodynamics. Phys. Rev. A {\bf 90}, 052110 (2014)

\bibitem{rastqip15}
Rastegin, A.E.: Fine-grained uncertainty relations for several quantum measurements. Quantum Inf. Process. {\bf 14}, 783--800 (2015)

\bibitem{prtv11}
Partovi, M.H.: Majorization formulation of uncertainty in quantum mechanics. Phys. Rev. A {\bf 84}, 052117 (2011)

\bibitem{prz13}
Pucha{\l}a, Z., Rudnicki, {\L}., \.{Z}yczkowski, K.: Majorization
entropic uncertainty relations. J. Phys. A: Math. Theor. {\bf 46},
272002 (2013)

\bibitem{fgg13}
Friedland, S., Gheorghiu, V., Gour, G.: Universal uncertainty relations, Phys. Rev. Lett. {\bf 111}, 230401 (2013)

\bibitem{rpz14}
Rudnicki, {\L}., Pucha{\l}a, Z., \.{Z}yczkowski, K.: Strong majorization entropic uncertainty relations. Phys. Rev. A {\bf 89}, 052115 (2014)

\bibitem{rud15}
Rudnicki, {\L}.: Majorization approach to entropic uncertainty relations for coarse-grained observables. Phys. Rev. A {\bf 91}, 032123 (2015)

\bibitem{povmkz16}
Rastegin, A.E., \.{Z}yczkowski, K.: Majorization entropic
uncertainty relations for quantum operations. J. Phys. A: Math.
Theor. {\bf 49}, 355301 (2016)

\bibitem{hj1990}
Horn, R.A., Johnson, C.R.: Matrix Analysis. Cambridge University Press, Cambridge (1990)

\bibitem{watrous1}
Watrous J.: Theory of Quantum Information, a draft of book. University of Waterloo, Waterloo (2017) \\
http://www.cs.uwaterloo.ca/{\textasciitilde}watrous/TQI/

\bibitem{renyi61}
R\'{e}nyi, A.: On measures of entropy and information. In: Neyman,
J. (ed.) Proceedings of 4th Berkeley Symposium on Mathematical
Statistics and Probability, vol. 1, pp. 547--561. University of
California Press, Berkeley (1961)

\bibitem{ja04}
Jizba, P., Arimitsu, T.: The world according to R\'enyi: thermodynamics of multifractal systems. Ann. Phys. {\bf 312}, 17--59 (2004)

\bibitem{bengtsson}
Bengtsson, I., \.{Z}yczkowski, K.: Geometry of Quantum States: An
Introduction to Quantum Entanglement. Cambridge University Press,
Cambridge (2006)

\bibitem{ben78}
Ben-Bassat, M., Raviv, J.: R\'{e}nyi's entropy and error probability. IEEE Trans. Inf. Theory {\bf 24}, 324--331 (1978)

\bibitem{tsallis}
Tsallis, C.: Possible generalization of Boltzmann--Gibbs statistics. J. Stat. Phys. {\bf 52}, 479--487 (1988)

\bibitem{bhatia97}
Bhatia, R.: Matrix Analysis. Springer, Berlin (1997)

\bibitem{bz10}
Durt, T., Englert, B.-G., Bengtsson, I., \.{Z}yczkowski, K.: On mutually unbiased bases. Int. J. Quantum Inf. {\bf 8}, 535--640 (2010)

\bibitem{wf89}
Wootters, W.K., Fields, B.D.: Optimal state-determination by mutually unbiased measurements. Ann. Phys. {\bf 191}, 363--381 (1989)

\bibitem{kr04}
Klappenecker, A., R\"{o}tteler, M.: Constructions of mutually
unbiased bases. In: Finite Fields and Applications, Lecture Notes
in Computer Science, vol. 2948, 137--144. Springer, Berlin (2004)

\bibitem{rbsc04}
Renes, J.M., Blume-Kohout, R., Scott, A.J., Caves, C.M.: Symmetric
informationally complete quantum measurements. J. Math. Phys. {\bf
45}, 2171--2180 (2004)

\bibitem{appl2005}
Appleby, D.M.: Symmetric informationally complete-positive
operator valued measures and the extended Clifford group. J. Math.
Phys. {\bf 46}, 052107 (2005)

\bibitem{grassl10}
Scott, A.J., Grassl, M.: Symmetric informationally complete
positive-operator-valued measures: a new computer study. J. Math.
Phys., {\bf 51}, 042203 (2010)

\bibitem{adf07}
Appleby, D.M., Dang, H.B., Fuchs, C.A.: Symmetric informationally-complete quantum states as analogues to
orthonormal bases and minimum-uncertainty states. arXiv:0707.2071 [quant-ph] (2007)

\bibitem{ruskai09}
Ruskai, M.B.: Some connections between frames, mutually unbiased
bases, and POVM's in quantum information theory. Acta Appl. Math.
{\bf 108}, 709--719 (2009)

\bibitem{kagour}
Kalev, A., Gour, G.: Mutually unbiased measurements in finite dimensions. New J. Phys. {\bf 16}, 053038 (2014)

\bibitem{kgour13}
Gour, G., Kalev, A.: Construction of all general symmetric informationally complete measurements. J. Phys. A: Math. Theor. {\bf 47}, 335302 (2014)

\bibitem{riesz27}
Riesz, M.: Sur les maxima des forms bilin\'{e}aires et sur les fonctionnelles lin\'{e}aires. Acta Math. {\bf 49}, 465--497 (1927)

\bibitem{rast104}
Rastegin, A.E.: Entropic uncertainty relations for extremal unravelings of super-operators. J. Phys. A: Math. Theor. {\bf 44}, 095303 (2011)

\bibitem{hall97}
Hall, M.J.W.: Quantum information and correlation bounds. Phys. Rev. A {\bf 55}, 100--113 (1997)

\bibitem{rastmub}
Rastegin, A.E.: Uncertainty relations for MUBs and SIC-POVMs in terms of generalized entropies. Eur. Phys. J. D {\bf 67}, 269 (2013)

\bibitem{remum15}
Rastegin, A.E.: On uncertainty relations and entanglement
detection with mutually unbiased measurements. Open Sys. Inf. Dyn.
{\bf 22}, 1550005 (2015)

\bibitem{molm09}
Wu, S., Yu, S., M{\o}lmer, K.: Entropic uncertainty relation for mutually unbiased bases. Phys. Rev. A {\bf 79}, 022104 (2009)

\bibitem{shbah12}
Spengler, C., Huber, M., Brierley, S., Adaktylos, T., Hiesmayr,
B.C.: Entanglement detection via mutually unbiased bases. Phys.
Rev. A {\bf 86}, 022311 (2012)

\bibitem{fei14}
Chen, B., Ma, T., Fei, S.-M.: Entanglement detection using mutually unbiased measurements. Phys. Rev. A {\bf 89}, 064302 (2014)

\bibitem{rastsic}
Rastegin, A.E.: Notes on general SIC-POVMs. Phys. Scr. {\bf 89}, 085101 (2014)

\bibitem{fei15}
Chen, B., Ma, T., Fei, S.-M.: General SIC measurement-based entanglement detection. Quantum Inf. Process. {\bf 14}, 2281--2290 (2015)

\bibitem{werner89}
Werner, R.F.: Quantum states with Einstein--Podolsky--Rosen
correlations admitting a hidden-variable model. Phys. Rev. A {\bf
40}, 4277--4281 (1989)

\bibitem{zhsl98}
\.{Z}yczkowski, K., Horodecki, P., Sanpera, A., Lewenstein, M.: Volume of the set of separable states. Phys. Rev. A {\bf 58}, 883--892 (1998)

\bibitem{dcls2017}
Das, S., Chanda, T., Lewenstein, M., Sanpera, A., Sen(De), A.,
Sen, U.: The separability versus entanglement problem.
arXiv:1701.02187 [quant-ph] (2017)

\bibitem{wtstd09}
Walborn, S.P., Taketani, B.G., Salles, A., Toscano, F., de Matos
Filho, R.L.: Entropic entanglement criteria for continuous
variables. Phys. Rev. Lett. {\bf 103}, 160505 (2009)

\bibitem{stw11}
Saboia, A., Toscano, F., Walborn, S.P.: Family of
continuous-variable entanglement criteria using general entropy
functions. Phys. Rev. A {\bf 83}, 032307 (2011)

\bibitem{rastpra17}
Rastegin, A.E.: R\'{e}nyi formulation of entanglement criteria for continuous variables. Phys. Rev. A {\bf 95}, 042334 (2017)

\bibitem{IBB06}
Bia{\l}ynicki-Birula, I.: Formulation of the uncertainty relations in terms of the R\'{e}nyi entropies. Phys. Rev. A {\bf 74}, 052101 (2006)

\bibitem{rast10r}
Rastegin, R.E.: R\'{e}nyi formulation of the entropic uncertainty principle for POVMs. J. Phys. A: Math. Theor. {\bf 43}, 155302 (2010)

\bibitem{ccyz12}
Coles, P.J., Colbeck, R., Yu, L., Zwolak, M.: Uncertainty relations from simple entropic properties. Phys. Rev. Lett. {\bf 108}, 210405 (2012)

\bibitem{popescu98}
Linden, N., Popescu, S.: On multi-particle entanglement. Fortschr. Phys. {\bf 46}, 567--578 (1998)

\bibitem{ibkz16}
Bengtsson, I., \.{Z}yczkowski, K.: A brief introduction to multipartite entanglement. arXiv:1612.07747 [quant-ph] (2016)

\bibitem{toth06}
T\'{o}th, G., G\"{u}hne, O.: Detection of multipartite entanglement with two-body correlations. Appl. Phys. B {\bf 82}, 237--241 (2006)

\bibitem{huangy12}
Huang, Y., Qiu, D.W.: Concurrence vectors of multipartite states based on coefficient matrices. Quantum
Inf. Process. {\bf 11}, 235--254 (2012)

\bibitem{huber13}
Spengler, C., Huber, M., Gabriel, A., Hiesmayr, B.C.: Examining the dimensionality of genuine multipartite
entanglement. Quantum Inf. Process. {\bf 12}, 269--278 (2013)

\bibitem{zhao15}
Zhao, C., Yang, G., Hung, W.N.N., Li, X.: A multipartite
entanglement measure based on coefficient matrices. Quantum Inf.
Process. {\bf 14}, 2861--2881 (2015)

\bibitem{rubin2000}
Pittenger, A.O., Rubin, M.N.: Note on separability of the Werner states in arbitrary dimensions. Opt. Commun. {\bf 179}, 447--449 (2000)

\end{thebibliography}
\end{document}